\documentstyle[preprint,aps]{revtex}	% PREPRINT STYLE
\begin{document}
\draft
%
%%%%%%%%%%%%%%%%%%%%% TITLE PAGE %%%%%%%%%%%%%%%%%%%%%%%%%%%%%%%%%%%%%%%%%%%%%%
%
\preprint{$
\begin{array}{l}
\mbox{FSU--HEP--941010}\\[-3mm]
\mbox{UCD--94--22}\\[-3mm]
\mbox{October~1994} \\   %[1cm]
\end{array}
$}
\title{$WZ$ Production at Hadron Colliders:  Effects of
Non-standard $WWZ$ Couplings and QCD Corrections}
\vskip -2.mm
\author{U.~Baur}
\address{
Department of Physics, Florida State University, Tallahassee, FL 32306,
USA\\[-1.mm]
and\\[-1.mm]
Department of Physics, SUNY at Buffalo, Buffalo, NY 14260, USA}
\vskip -2.mm
\author{T.~Han and J.~Ohnemus}
\address{
Department of Physics, University of California, Davis, CA 95616, USA}
\maketitle
\vskip -3.mm
\begin{abstract}
\vskip -8.mm
\baselineskip15.pt  %to keep abstract on 1 page
The process $p\,p\hskip-7pt\hbox{$^{^{(\!-\!)}}$} \rightarrow
W^{\pm}Z + X \rightarrow \ell^\pm_1 \nu_1 \ell_2^+ \ell_2^- + X$
is calculated to
${\cal O}(\alpha_s)$ for general $C$ and $P$ conserving $WWZ$ couplings.
At the Tevatron center of mass energy, the QCD corrections to $WZ$
production are modest. At Large Hadron Collider (LHC)
energies, the inclusive QCD corrections are large, but
can be reduced significantly if a jet veto is imposed.
Sensitivity limits for the anomalous $WWZ$ couplings are derived from
the next-to-leading order $Z$ boson transverse momentum
distribution for Tevatron and LHC energies. Unless a jet veto is
imposed, next-to-leading order QCD corrections decrease the sensitivity to
anomalous $WWZ$ couplings considerably at LHC energies, but have little
influence at the Tevatron. We also
study, at next-to-leading order, rapidity correlations between
the $W$ and $Z$ decay products, and the $ZZ/WZ$ and $WZ/W\gamma$ cross
section ratios. These quantities are found to be useful tools in
searching for the approximate zero present in the Standard Model $WZ$
helicity amplitudes. The prospects for observing the approximate amplitude
zero at the Tevatron and the LHC are critically assessed.

\end{abstract}
\vskip .5in
\newpage
%
%%%%%%%%%%%%%%%%%% MAIN TEXT %%%%%%%%%%%%%%%%%%%%%%%%%%%%%%%%%%%%%%%%%%%%%%%
%
\narrowtext

\section{INTRODUCTION}

The electroweak Standard Model (SM) based on an
$\hbox{\rm SU(2)} \bigotimes \hbox{\rm U(1)}$ gauge theory
has been remarkably successful in describing contemporary high
energy physics experiments. The three vector boson couplings
predicted by this non-abelian gauge theory, however, remain largely untested.
The production of $WZ$ pairs at hadron colliders provides an
excellent opportunity to study the $WWZ$
vertex~\cite{FIRSTWZ,EHLQ,WILLEN,WWZ,UJH}.  In addition,
the reaction $p\,p\hskip-7pt\hbox{$^{^{(\!-\!)}}$} \rightarrow
W^{\pm}Z$ is of interest due to the presence of an approximate zero in the
amplitude of the parton level subprocess $q_1\bar q_2\rightarrow W^\pm
Z$~\cite{UJH} in the SM, which is similar in nature to the well-known
radiation zero in the reaction
$p\,p\hskip-7pt\hbox{$^{^{(\!-\!)}}$} \rightarrow
W^{\pm}\gamma$~\cite{RAZ}. In the SM, the $WWZ$ vertex is completely
fixed by the $\hbox{\rm SU(2)} \bigotimes \hbox{\rm U(1)}$ gauge
structure of the electroweak sector. A measurement of the $WWZ$
vertex thus provides a stringent test of the SM.

$WZ$ production at hadron colliders has recently
received attention due to the observation of a clean
$W^+Z\rightarrow e^+\nu_e e^+e^-$ candidate event by the CDF
Collaboration~\cite{GLAS}. Double leptonic $WZ$
decays are relatively background free and therefore provide an
excellent testing ground for anomalous $WWZ$ couplings.
With an integrated luminosity of the order of
1~fb$^{-1}$, which is envisioned for the Main Injector era~\cite{GPJ},
a sufficient number of events should be
available to commence a detailed investigation of the $WWZ$ vertex in
the $W^\pm Z \to \ell^\pm_1 \nu_1 \ell_2^+\ell_2^-$ channel
($\ell_1,\,\ell_2=e,\,\mu$). The
prospects for a precise measurement of the $WWZ$ couplings in this
channel would further improve if integrated luminosities on the
order of 10~fb$^{-1}$ could be achieved (a luminosity upgraded
Tevatron will henceforth be denoted by TeV*) and/or
the energy of the Tevatron could be doubled to $\sqrt{s}=3.5-4$~TeV
(an energy upgraded Tevatron will henceforth be referred to
as the DiTevatron)~\cite{GPJ}. At the Large Hadron Collider (LHC, $pp$
collisions at $\sqrt{s}=14$~TeV~\cite{LHC}), it should be possible to
determine the $WWZ$ couplings with high precision~\cite{WILLEN}.

In contrast to low energy data and high precision measurements at the $Z$
peak, collider experiments offer the possibility of a direct, and
essentially model independent, determination of the three vector boson
vertices. Hadronic production of $WZ$ pairs was first calculated in
Ref.~\cite{FIRSTWZ}. The ${\cal O}(\alpha_s)$ QCD corrections to the
reaction $p\,p\hskip-7pt\hbox{$^{^{(\!-\!)}}$} \rightarrow W^{\pm}Z$
were first evaluated in Refs.~\cite{WZ} and \cite{FRIX}.
Studies on the potential for probing the $WWZ$
vertex have been performed in Refs.~\cite{WILLEN} and~\cite{WWZ}.

Previous studies on probing the $WWZ$ vertex via hadronic $WZ$
production have been based on leading-order (LO)
calculations \cite{WILLEN,WWZ}.  In general, the
inclusion of anomalous couplings at the $WWZ$ vertex yields
enhancements in the $WZ$ cross section, especially at large values of
the $W$ or $Z$ boson transverse momentum, $p_T^{}(W)$ or $p_T^{}(Z)$,
and at large values of the $WZ$ invariant mass, $M_{WZ}$.
Next-to-leading-order (NLO) calculations of hadronic $WZ$
production have shown that the ${\cal O}(\alpha_s)$ corrections
are large in precisely these same regions \cite{WZ,FRIX}.
It is thus vital to include the NLO corrections when using
hadronic $WZ$ production to test the $WWZ$ vertex for anomalous couplings.

In this paper, we calculate hadronic $WZ$ production to ${\cal
O}(\alpha_s)$, including the most general, $C$ and $P$ conserving, anomalous
$WWZ$ couplings.  Our calculation also includes the leptonic decays
of the $W$ and $Z$ bosons in the narrow width approximation. Decay
channels where the $W$ or $Z$ boson decays hadronically are not
considered here. The calculation has been performed
using the Monte Carlo method for NLO calculations~\cite{NLOMC}.
With this method, it
is easy to calculate a variety of observables simultaneously and to
implement experimental acceptance cuts in the calculation. It is also
possible to compute the NLO QCD corrections for exclusive channels,
{\it e.g.}, $p\,p\hskip-7pt\hbox{$^{^{(\!-\!)}}$} \rightarrow WZ
+0$~jet. Apart from anomalous contributions to the $WWZ$ vertex we
assume the SM to be valid in our calculation. In particular, we assume
the couplings of the $W$ and $Z$ bosons to quarks and leptons
are as given by the SM. Section~II briefly summarizes the technical
details of our calculation.

The results of our numerical simulations are presented in Secs.~III and~IV.
In contrast to the SM contributions to the $q_1\bar q_2\to
WZ$ helicity amplitudes, terms
associated with non-standard $WWZ$ couplings grow with energy.
The $WZ$ invariant mass distribution, the cluster
transverse mass distribution, and the $Z$ boson transverse momentum
spectrum are therefore very sensitive to anomalous $WWZ$ couplings. In
Sec.~III, we focus on the impact QCD corrections have on these
distributions, and the sensitivity limits for the anomalous $WWZ$
couplings which can be achieved at the Tevatron, DiTevatron, and LHC with
various integrated luminosities. At
LHC energies, the inclusive NLO QCD corrections in the SM are found to
be very large at high $p_T^{}(W)$ or $p_T^{}(Z)$, and have a
non-negligible influence on the sensitivity bounds which can be achieved
for anomalous $WWZ$ couplings.  The large QCD corrections are caused by
the combined effects of destructive interference in the Born subprocess,
a log-squared enhancement factor in the $q_1 g \rightarrow W Z q_2$
partonic cross section at high
transverse momentum \cite{FRIX}, and the large quark-gluon luminosity
at supercollider energies. At the Tevatron, on the other hand, the
${\cal O}(\alpha_s)$ QCD corrections are modest and
sensitivities are only slightly affected by the QCD corrections.
In Sec.~III, we also show that the size of the QCD corrections at high
$p_T^{}(Z)$ or $p_T^{}(W)$ can be significantly reduced, and a
significant fraction of the sensitivity lost at LHC energies can be
regained, if a jet veto
is imposed, {\it i.e.}, if the $WZ+0$~jet exclusive channel is considered.
We also find that the residual dependence of the NLO
$WZ+0$~jet cross section on the factorization scale
$Q^2$ is significantly smaller than that of the ${\cal O}(\alpha_s)$
cross section for the inclusive reaction $p\,p\hskip-7pt
\hbox{$^{^{(\!-\!)}}$} \rightarrow WZ+X$.

In Sec.~IV, we study rapidity correlations between
the $W$ and $Z$ decay products, the $ZZ/WZ$ and $WZ/W\gamma$ cross
section ratios, and how these quantities are affected by QCD corrections.
In the SM at tree level, the process $q_1\bar q_2
\rightarrow WZ$ exhibits an approximate amplitude zero at
$ \cos\Theta^* \approx  {1\over 3}\tan^2\theta_{\rm W}\approx  0.1$
($\cos\Theta^* \approx -{1\over 3}\tan^2\theta_{\rm W}\approx -0.1$)
for $u\bar d\rightarrow W^+Z$ ($d\bar u\rightarrow W^-Z$)~\cite{UJH}
which is similar in nature to the well-known radiation zero in $W\gamma$
production in hadronic
collisions. Here $\Theta^*$ is the $Z$ boson scattering angle in the
parton center of mass frame relative to the quark direction and
$\theta_{\rm W}$ is the weak mixing angle.
The radiation zero in $W\gamma$ production can be observed rather
easily at the Tevatron in the
distribution of the rapidity difference $\Delta y(\gamma,\ell)
=y(\gamma)-y(\ell)$ of the photon and the charged lepton which
originates from the $W$ decay, $W\rightarrow\ell\nu$~\cite{BEL}. In the
SM, the $\Delta y(\gamma,\ell)$ distribution exhibits a pronounced dip
at $\Delta y(\gamma,\ell)\approx\mp 0.3$ in $W^\pm\gamma$ production, which
originates from the radiation zero. We find that, at leading order,
the approximate amplitude zero in $p\,p\hskip-7pt
\hbox{$^{^{(\!-\!)}}$} \rightarrow W^\pm
Z\rightarrow\ell_1^\pm\nu_1\ell_2^+\ell_2^-$ leads to a
dip in the corresponding $\Delta y(Z,\ell_1)= y(Z) - y(\ell_1)$
distribution which is located at $\Delta y(Z,\ell_1)\approx\pm 0.5$
($=0$) in $p\bar p$ ($pp$) collisions. NLO QCD corrections tend
to fill in the dip; at
LHC energies they obscure the signal of
the approximate amplitude zero almost completely, unless a jet
veto is imposed.

Alternatively, cross section ratios can be considered.
Analogous to the $Z\gamma/W\gamma$ cross section ratio~\cite{BEO}, the
$ZZ/WZ$ cross section ratio as a function of the minimum transverse
momentum of the $Z$ boson is found to reflect the approximate
amplitude zero. The ratio of the $WZ$ to $W\gamma$ cross sections, on the
other hand, measures
the relative strength of the radiation zero in $W\gamma$ production and
the approximate zero in $q_1\bar q_2\to WZ$. QCD corrections
significantly affect the ratio of $ZZ$ to $WZ$ cross sections, but
largely cancel in the $WZ/W\gamma$ cross section ratio. Although rapidity
correlations and cross section ratios are useful tools in searching for
the approximate amplitude zero in $WZ$ production, it will not be easy
to establish the effect at the Tevatron or LHC, due to the limited
number of $W^\pm Z\to\ell_1^\pm\nu_1\ell_2^+\ell_2^-$ events expected.
Our conclusions, finally, are given in Sec.~V.

\section{CALCULATIONAL TOOLS}

Our calculation generalizes the results of Ref.~\cite{NEWJO} to
arbitrary $C$ and $P$ conserving $WWZ$ couplings. It is carried out
using a combination of analytic and Monte Carlo integration techniques.
Details of the method can be found in Ref.~\cite{NLOMC}.
The calculation is done using the narrow width approximation for
the leptonically decaying $W$ and $Z$ bosons.
In this approximation it is particularly easy to extend
the NLO calculation of $WZ$ production with on-shell $W$ and $Z$ bosons
to include the leptonic $W$ and $Z$ decays.
Furthermore, non-resonant Feynman diagrams such as
$d \bar u \to W^{-*} \to   e^- \bar \nu_eZ$ followed by $Z \to \mu^+
\mu^-$ contribute negligibly in this limit, and thus can be ignored.

\subsection{Summary of ${\cal O}(\alpha_s)$ $WZ$ Production including
$W$ and $Z$ decays}

The NLO calculation of $WZ$ production includes contributions from the
square of the Born graphs, the interference
between the Born graphs and the virtual one-loop diagrams, and the square
of the real emission graphs. The basic idea of the method employed here
is to isolate the
soft and collinear singularities associated with the real emission
subprocesses by partitioning phase space into soft, collinear, and
finite regions.  This is done by introducing theoretical soft and
collinear cutoff parameters, $\delta_s$ and $\delta_c$.  Using
dimensional regularization, the soft and collinear
singularities are exposed as poles in  $\epsilon$ (the number of
space-time dimensions is $N = 4 - 2\epsilon$ with $\epsilon$ a small
number). The infrared singularities from the soft and virtual
contributions are then explicitly canceled while the collinear
singularities are factorized and absorbed into the definition of the
parton distribution functions. The remaining contributions are finite
and can be  evaluated in four dimensions.  The Monte Carlo program thus
generates $n$-body (for the Born and virtual contributions) and
$(n+1)$-body (for the real emission contributions) final state events.
The $n$- and $(n+1)$-body contributions both depend on the cutoff
parameters $\delta_s$ and $\delta_c$, however, when these contributions
are added together to form a suitably inclusive observable, all
dependence on the cutoff parameters cancels.
The numerical results presented in this paper are insensitive to
variations of the cutoff parameters.

Except for the virtual contribution,
the ${\cal O}(\alpha_s)$ corrections are all proportional to the Born cross
section.  It is easy to incorporate the decays $W \to \ell_1 \nu_1$ and
$Z\to\ell_2^+\ell_2^-$ into those terms which
are proportional to the Born cross section; one simply replaces
$d\hat\sigma^{\hbox{\scriptsize Born}} (q_1 \bar q_2 \to WZ)$ with
$d\hat\sigma^{\hbox{\scriptsize Born}}
(q_1 \bar q_2 \to WZ\to \ell_1 \nu_1\ell_2^+\ell_2^-)$ in the relevant
formulae. When working at the amplitude level, the $W$ and $Z$ decays
are trivial to implement; the vector boson polarization vector
$\epsilon^V_\mu (k)$, $V=W,\, Z$, is simply replaced by the respective
decay current $J^V_\mu (k)$ in the
amplitude.  Details of the amplitude level calculations for the Born
and real emission subprocesses can be found in Ref.~\cite{VVJET}. For
$\ell_1=\ell_2$ the amplitudes in principle should be antisymmetrized.
Since the leptons originating from the decay of the $W$ and $Z$ bosons
are usually
well separated, effects from the antisymmetrization of the
amplitudes are expected to be very small and hence are ignored here.

The only term in which it is more difficult to incorporate the $W$ and
$Z$ decays is
the virtual contribution.  Rather than undertake the non-trivial task of
recalculating the virtual correction term for the case of
leptonically decaying  weak bosons, we have instead opted to use the
virtual correction for real on-shell
$W$ and $Z$ bosons which we subsequently decay ignoring spin correlations.
When spin correlations are ignored, the spin summed squared matrix element
factorizes into separate production and decay squared matrix elements.

Neglecting spin correlations slightly modifies the shapes of the
angular distributions of the final state leptons, but
does not alter the total cross section as long as
no angular cuts ({\it e.g.}, rapidity cuts)
are imposed on the final state leptons. For realistic
rapidity cuts, cross sections are changed by typically 10\% if spin
correlations are neglected.
Since the size of the finite virtual correction is less than
$\sim 10\%$ the size of the Born cross section, the overall effect of
neglecting the spin correlations in the finite virtual correction is expected
to be negligible compared to the combined $10\% \sim 20\%$ uncertainty
from the parton distribution functions, the choice of the scale $Q^2$,
and higher order QCD corrections.

\subsection{Incorporation of Anomalous $WWZ$ Couplings}

The $WWZ$ vertex is uniquely determined in the SM by
$\hbox{\rm SU(2)} \bigotimes \hbox{\rm U(1)}$
gauge invariance.  In $WZ$ production the $W$ and $Z$ bosons
couple to essentially massless fermions, which
insures that effectively $\partial_\mu W^\mu=0$ and
$\partial_\mu Z^\mu=0$. These conditions
together with Lorentz invariance and conservation of $C$ and $P$,
allow three free parameters, $g_1, \kappa$, and $\lambda$,
in the $WWZ$ vertex. The most general Lorentz, $C$, and $P$ invariant
vertex is described by the effective Lagrangian~\cite{LAGRANGIAN}
\begin{eqnarray}
\noalign{\vskip 5pt}
{\cal L}_{WWZ} &=& -i \, e \, {\rm cot} \theta_{\rm W}
\Biggl[ g_1 \bigl( W_{\mu\nu}^{\dagger} W^{\mu} Z^{\nu}
                  -W_{\mu}^{\dagger} Z_{\nu} W^{\mu\nu} \bigr)
+ \kappa W_{\mu}^{\dagger} W_{\nu} Z^{\mu\nu}
+ {\lambda \over M_W^2} W_{\lambda \mu}^{\dagger} W^{\mu}_{\nu}
Z^{\nu\lambda} \Biggr] \>,
\label{EQ:LAGRANGE}
\end{eqnarray}
\narrowtext
where $W^{\mu}$ and $Z^{\mu}$ are the $W^-$ and $Z$ fields, respectively,
$W_{\mu\nu} = \partial_{\mu}W_{\nu} - \partial_{\nu}W_{\mu}$, and
$Z_{\mu\nu} = \partial_{\mu}Z_{\nu} - \partial_{\nu}Z_{\mu}$.
At tree level in the SM, $g_1=1$, $\kappa = 1$, and $\lambda = 0$.
All higher dimensional operators are obtained by replacing $V^\mu$ with
$(\partial^2)^m V^\mu$, $V=W,\,Z$, where $m$ is an arbitrary positive
integer, in the terms proportional to $\Delta g_1 = g_1 -1$,
$\Delta\kappa=\kappa-1$, and $\lambda$.
These operators form a complete set and can be summed by
replacing $\Delta g_1$, $\Delta\kappa$,
and $\lambda$ by momentum dependent form factors. All
details are contained in the specific functional form of the form factor
and its scale $\Lambda_{FF}$. The form factor nature of $\Delta g_1$,
$\Delta\kappa$, and
$\lambda$ will be discussed in more detail later in this section.

Following the standard notation of Ref.~\cite{LAGRANGIAN}, we have
chosen, without loss of generality, the
$W$ boson mass, $M_W$, as the energy scale in the denominator of the term
proportional to $\lambda$ in Eq.~(\ref{EQ:LAGRANGE}). If a different
mass scale, ${\tt M}$, had been used, then all of our subsequent
results could be obtained by scaling $\lambda$ by a factor
${\tt M}^2/M_W^2$.

At present, the $WWZ$ coupling constants are only weakly constrained
experimentally. The CDF Collaboration recently presented preliminary
95\% confidence level (CL) limits on
$\Delta\kappa$ and $\lambda$ from a search performed in the $p\bar p\to
WZ \to jj \ell^+\ell^-$ ($\ell=e,\,\mu$) channel at large di-jet
transverse momenta, $p_T^{}(jj)>100$~GeV~\cite{GLAS}:
\begin{eqnarray}
\noalign{\vskip 5pt}
-8.6<\Delta\kappa<9.0~~({\rm for}~\lambda=0)
\>, \hskip 1.cm
-1.7<\lambda<1.7~~({\rm for}~\Delta\kappa=0) \>.
\end{eqnarray}
Assuming SM $WW\gamma$ couplings, CDF also obtained a limit on
$\Delta\kappa$ from the reactions $p\bar p\to W^+W^-,\,W^\pm Z\to\ell^\pm
\nu jj$ with $p_T^{}(jj)>130$~GeV~\cite{GLAS,TH}:
\begin{eqnarray}
\noalign{\vskip 5pt}
-1.3<\Delta\kappa<1.4~~~~({\rm for}~\lambda=0) \>.
\end{eqnarray}
To derive these limits, a dipole form factor with scale $\Lambda_{FF}=1.
5$~TeV was assumed (see below), however, the experimental bounds
are quite insensitive to the value of $\Lambda_{FF}$.
Although bounds on these couplings can also be extracted from
low energy data and high precision measurements at the $Z$ pole, there are
ambiguities and model dependencies in the results \cite{DE,BL,HISZ}.
{}From loop contributions to rare meson decays such as $K_L \to
\mu^+\mu^-$~\cite{HE} or $B\to K^{(*)}\mu^+\mu^-$~\cite{BAL},
$\epsilon'/\epsilon$~\cite{HMK}, and the $Z\to b\bar b$ width~\cite{NOV},
one estimates limits for the non-standard $WWZ$ couplings of
${\cal O}(1-10)$. No rigorous bounds can be
obtained from LEP~I data if correlations between different
contributions to the anomalous couplings are  taken fully into account.
In contrast, invoking a ``naturalness'' argument
based on chiral perturbation theory~\cite{FLS,BDV},
one expects deviations from the SM of ${\cal O}(10^{-2})$
or less for $g_1$, $\kappa$, and $\lambda$.

If $C$ or $P$ violating $WWZ$ couplings are allowed, four additional
free parameters, $g_4$, $g_5$, $\tilde\kappa$, and $\tilde\lambda$
appear in the effective Lagrangian. For simplicity, these couplings are
not considered in this paper.

The Feynman rule for the $WWZ$ vertex factor
corresponding to the Lagrangian in Eq.~(\ref{EQ:LAGRANGE}) is
\begin{eqnarray}
-i \, g_{WWZ} \, Q_W \, \Gamma_{\beta \mu \nu}^{} (k, k_1, k_2) =
-i \, g_{WWZ} \, Q_W \,
\biggl[
\Gamma_{\beta \mu \nu}^{\hbox{\scriptsize SM}} (k, k_1, k_2)
+ \Gamma_{\beta \mu \nu}^{\hbox{\scriptsize NSM}} (k, k_1, k_2) \biggr] \>,
\label{EQ:NSMCOUPLINGS}
\end{eqnarray}
where the labeling conventions for the four-momenta and Lorentz
indices are defined by Fig.~\ref{FIG:VERTEX}, $g_{WWZ}=e\,
\cot\theta_{\rm W}$ is the $WWZ$ coupling strength, $Q_W$ is the
electric charge of the $W$ boson in units
of the proton charge $e$, and the factors
$\Gamma^{\hbox{\scriptsize SM}}$ and $\Gamma^{\hbox{\scriptsize NSM}}$
are the SM and non-standard model vertex factors:
\begin{eqnarray}
\noalign{\vskip 5pt}
\Gamma_{\beta \mu \nu}^{\hbox{\scriptsize SM}} (k, k_1, k_2) =
&\phantom{+}& (k_1 - k_2)_{\beta} \, g_{\nu \mu} + 2 \,
 k_{\mu} \, g_{\beta \nu} - 2 \, k_{\nu} \, g_{\beta \mu} \>, \\
\noalign{\vskip 5pt}
\Gamma_{\beta \mu \nu}^{\hbox{\scriptsize NSM}} (k, k_1, k_2)
= &\phantom{+}&  {1\over 2} \,
\left( \Delta g_1 + \Delta\kappa + \lambda \,{k^2 \over M_W^2} \right)
        (k_1 - k_2)_{\beta} \, g_{\nu \mu} \\
\noalign{\vskip 5pt}
&-& {\lambda \over M_W^2} \, (k_1 - k_2)_{\beta} \, k_{\nu} \, k_{\mu}
 + (\Delta g_1 + \Delta\kappa + \lambda) \, k_{\mu} \, g_{\beta \nu}
\nonumber \\
\noalign{\vskip 5pt}
&-& \left( 2 \Delta g_1 + \lambda\, {M_Z^2 \over M_W^2} \right)
\, k_{\nu} \, g_{\beta \mu} \>. \nonumber
\end{eqnarray}
\narrowtext
The non-standard model vertex factor is written here in terms of
$\Delta g_1 = g_1 - 1$, $\Delta \kappa = \kappa - 1$,
and $\lambda$, which all vanish in the SM.

It is straightforward to include the non-standard model couplings in
the amplitude level calculations.  Using the computer algebra program
FORM~\cite{FORM}, we have computed the $q_1 \bar q_2 \to W Z$
virtual correction with the modified vertex
factor of Eq.~(\ref{EQ:NSMCOUPLINGS}), however,
the resulting expression is too lengthy to present here. The
non-standard $WWZ$ couplings of Eq.~(\ref{EQ:LAGRANGE}) do not
destroy the renormalizability of QCD. Thus the infrared singularities
from the soft and virtual contributions are explicitly canceled, and the
collinear singularities are factorized and absorbed into the definition
of the parton distribution functions, exactly as in the SM case.

The anomalous couplings can not be simply inserted into the vertex
factor as constants because this would violate $S$-matrix unitarity.
Tree level unitarity uniquely restricts the $WWZ$ couplings to their
SM gauge theory values at asymptotically high energies~\cite{CORNWALL}.
This implies that any deviation of $\Delta g_1$, $\Delta \kappa$,
or $\lambda$ from the SM expectation has to be described by a form
factor $\Delta g_1(M_{WZ}^2, p_W^2, p_Z^2)$,
$\Delta \kappa(M_{WZ}^2, p_W^2, p_Z^2)$, or
$\lambda(M_{WZ}^2, p_W^2, p_Z^2)$
which vanishes when either the square of the $WZ$ invariant mass,
$M_{WZ}^2$, or the
square of the four-momentum of the final state $W$ or $Z$ ($p_W^2$ or
$p_Z^2)$ becomes large.  In $WZ$ production
$p_W^2 \approx M_W^2$ and $p_Z^2 \approx M_Z^2$
even when the finite $W$ and $Z$ widths are taken into account.
However, large values of $M_{WZ}^2$ will be probed at future hadron
colliders like the LHC and the $M_{WZ}^2$ dependence
of the anomalous couplings has
to be included in order to avoid unphysical results which would violate
unitarity.  Consequently, the anomalous couplings are introduced
via form factors~\cite{FORMF,BAURZEP}
\begin{eqnarray}
\noalign{\vskip 5pt}
\Delta g_1(M_{WZ}^2, p_W^2 = M_W^2, p_Z^2 = M_Z^2) \> &=& \>
{\Delta g_1^0 \over (1 + M_{WZ}^2/\Lambda_{FF}^2)^n } \>,
\label{EQ:GONEFORM} \\
\noalign{\vskip 5pt}
\Delta \kappa(M_{WZ}^2, p_W^2 = M_W^2, p_Z^2 = M_Z^2) \> &=& \>
{\Delta \kappa^0 \over (1 + M_{WZ}^2/\Lambda_{FF}^2)^n } \>,
\label{EQ:KAPPAFORM} \\
\noalign{\vskip 5pt}
\lambda(M_{WZ}^2, p_W^2 = M_W^2, p_Z^2 = M_Z^2) \> &=& \>
{\lambda^0 \over (1 + M_{WZ}^2/\Lambda_{FF}^2)^n } \>,
\label{EQ:LAMBDAFORM}
\end{eqnarray}
where $\Delta g_1^0$, $\Delta \kappa^0$,
and $\lambda^0$ are the form factor values at low energies and
$\Lambda_{FF}$ represents the scale at which new physics becomes
important in the weak boson sector, {\it e.g.}, due to new resonances
or composite structures of the $W$ and $Z$ bosons.
In order to guarantee unitarity, it is necessary to have
$n>1/2$ for $\Delta\kappa$ and $n>1$
for $\Delta g_1$ and $\lambda$. For the numerical results presented
here, we use a dipole
form factor ($n=2$) with a scale $\Lambda_{FF} = 1$~TeV, unless
explicitly stated otherwise. The exponent $n=2$
is chosen in order to suppress $WZ$ production at energies
$\sqrt{\hat s} > \Lambda_{FF} \gg M_W,\,M_Z$, where novel phenomena
like resonance
or multiple weak boson production are expected to become important.

Form factors are usually not introduced if an ansatz based on chiral
perturbation theory is used.  In the framework of chiral perturbation
theory, the effective Lagrangian describing the anomalous vector boson
self-interactions breaks down at center of mass energies above a few
TeV~\cite{FLS,BDV} (typically $4\pi v \sim 3$~TeV, where $v\approx 246$~GeV
is the Higgs field vacuum expectation value).
Consequently, one has to limit the center of mass energies to
values sufficiently below $4\pi v$ in this approach.

\section{QCD CORRECTIONS AND NON-STANDARD $WWZ$ COUPLINGS}

We shall now discuss the phenomenological implications of NLO QCD
corrections to $WZ$ production at the Tevatron
($p\bar p$ collisions at $\sqrt{s} = 1.8$~TeV) and the LHC ($pp$
collisions at $\sqrt{s} = 14$~TeV). We also consider a possible
upgrade~\cite{GPJ} of the Tevatron to $\sqrt{s}=3.5$~TeV (DiTevatron).
We first briefly describe
the input parameters, cuts, and the finite energy resolution smearing
used to simulate detector response. We then discuss in detail the impact
of NLO QCD corrections on the observability of non-standard $WWZ$
couplings in $WZ$ production at the Tevatron, DiTevatron, and LHC. To
simplify the discussion, we shall concentrate on $W^+Z$ production. In
$p\bar p$ collisions the rates for $W^+Z$ and $W^-Z$
production are equal. At $pp$ colliders, the $W^-Z$ cross section
is about 30\% smaller than that of $W^+Z$ production. Furthermore,
we shall only consider $W^+\to \ell_1^+ \nu_1$ and $Z\to \ell_2^+
\ell_2^-$ decays ($\ell_1,\,\ell_2 = e,\,\mu$).

\subsection{Input Parameters}

The numerical results presented here
were obtained using the two-loop expression for
$\alpha_s$. The QCD scale $\Lambda_{\hbox{\scriptsize QCD}}$
is specified for four
flavors of quarks by the choice of the parton distribution functions  and
is adjusted whenever a heavy quark threshold is crossed so that
$\alpha_s$ is a continuous function of $Q^2$. The heavy quark masses
were taken to be $m_b=5$~GeV and $m_t=150$~GeV, which is consistent with
the bound obtained by D\O, $m_t>131$~GeV~\cite{DTOP}, and the value
suggested by the current CDF data, $m_t=174\pm
10\,^{+13}_{-12}$~GeV~\cite{CDFTOP}.
Our results are insensitive to the value chosen for $m_t$.

The SM parameters used in the numerical simulations are $M_Z = 91.173$~GeV,
$M_W = 80.22$~GeV, $\alpha (M_W) =1/128$, and $\sin^2
\theta_{\hbox{\scriptsize W}} = 1 - (M_W^{}/M_Z^{})^2$. These values are
consistent with recent measurements at LEP, SLC, the CERN $p\bar p$
collider, and the Tevatron \cite{LEP,SLC,MT,MW}. The soft and collinear
cutoff parameters, as discussed in Sec.~IIA,
are fixed to $\delta_s = 10^{-2}$ and $\delta_c = 10^{-3}$. The parton
subprocesses have been summed over $u,d,s$, and $c$ quarks and the
Cabibbo mixing angle has been chosen such that $\cos^2 \theta_C =
0.95$. The leptonic branching ratios have been taken to be
$B(W \to \ell \nu) = 0.107$ and $B(Z \to \ell^+\ell^-) = 0.034$
and the total widths of the $W$ and $Z$ bosons are
$\Gamma_W = 2.12$~GeV and $\Gamma_Z = 2.487$~GeV.
Except where otherwise stated, a single scale
$Q^2=M^2_{WZ}$,  where $M_{WZ}$ is the invariant mass of the
$WZ$ pair, has been used for the renormalization scale $\mu^2$
and the factorization scale $M^2$.

In order to get consistent NLO results it is necessary to use parton
distribution functions which have been fit to next-to-leading order. In
our numerical simulations we have used
the Martin-Roberts-Stirling (MRS)~\cite{MRSPRIME} set S0$^\prime$
distributions
with $\Lambda_4 = 230$~MeV, which take into account the most recent
NMC~\cite{NMC} and CCFR~\cite{CCFR} data and are consistent with
measurements of the proton structure functions at HERA~\cite{HERA}.
For convenience, the MRS set S0$^\prime$
distributions have also been used for the LO calculations.

\subsection{Cuts}

The cuts imposed in our numerical simulations are motivated by
the finite acceptance of detectors.
The complete set of transverse momentum ($p_T^{}$), pseudorapidity ($\eta$),
and separation cuts can be summarized as follows.
\newpage
\begin{quasitable}
\begin{tabular}{cc}
Tevatron & LHC\\
\tableline
$p_{T}^{}(\ell)         > 20$~GeV  & $p_{T}^{}(\ell)         > 25$~GeV\\
$p\llap/_T^{}           > 20$~GeV  & $p\llap/_T^{}           > 50$~GeV\\
$|\eta(\ell)|           < 2.5$     & $|\eta(\ell)|           < 3.0$\\
$\Delta R(\ell,\ell)    > 0.4$     & $\Delta R(\ell,\ell)    > 0.4$\\
\end{tabular}
\end{quasitable}
Here, $\Delta R = [ (\Delta\phi)^2 + (\Delta\eta)^2]^{1/2}$ is the
separation in the pseudorapidity -- azimuthal angle plane. The $\Delta
R(\ell,\ell)$ cut is only imposed on leptons of equal electric charge.
It has only a small effect on the $WZ$ cross section.
For simplicity, identical cuts are imposed on final
state electrons and muons. The large missing transverse momentum
($p\llap/_T^{}$) cut at LHC
energies, which severely reduces the total $WZ$ cross section, has been
chosen to reduce potentially dangerous backgrounds from event
pileup~\cite{PILE}, $pp\to Zb\bar b\to\ell_1\nu_1\ell_2^+\ell_2^-+X$,
and processes where particles outside the rapidity range covered by the
detector contribute to the missing transverse momentum. Present
studies~\cite{ATLAS,CMS} indicate that these backgrounds
are under control for
$p\llap/_T^{}>50$~GeV. The total $W^+Z$ cross section within cuts in the
Born approximation at the Tevatron, DiTevatron, and LHC is 8.5~fb,
22.4~fb, and 25.9~fb, respectively. If the LHC is operated significantly
below the design luminosity of ${\cal
L}=10^{34}$~cm$^{-2}$~s$^{-1}$~\cite{LHC}, the background from event
pileup is less severe and
it may well be possible to lower the missing $p_T^{}$ threshold. If the
$p\llap/_T>50$~GeV cut is replaced by a $p\llap/_T>20$~GeV requirement,
the total LO $W^+Z$ cross section triples to 80.8~fb.

\subsection{Finite Energy Resolution Effects}

Uncertainties in the energy measurements of the charged leptons
in the detector are simulated in our calculation by Gaussian smearing
of the particle four-momentum vector with standard deviation $\sigma$.
For distributions which require a jet definition, {\it e.g.}, the $WZ +
1$~jet exclusive cross section, the jet four-momentum vector is also
smeared. The standard deviation $\sigma$
depends on the particle type and the detector. The numerical results
presented here for the Tevatron/DiTevatron and LHC center of mass energies
were made using $\sigma$ values based on the CDF~\cite{RCDF} and
ATLAS~\cite{ATLAS} specifications, respectively.

\subsection{Inclusive NLO Cross Sections}

The sensitivity of $WZ$ production to anomalous $WWZ$
couplings in the Born approximation was studied in detail in
Refs.~\cite{WILLEN} and \cite{WWZ}. The distributions of the $Z$ boson
transverse momentum, $p_{T}^{}(Z)$, and the $WZ$ invariant mass,
$M_{WZ}^{}$, were found to be sensitive to the anomalous couplings.
However, at hadron colliders the $WZ$ invariant mass cannot be determined
unambiguously because the neutrino from the $W$ decay is not observed.
If the transverse momentum of the neutrino is identified with the missing
transverse momentum of a given $WZ$ event, the unobserved longitudinal
neutrino momentum $p_L^{}(\nu)$ can be reconstructed, albeit with a
twofold ambiguity, by imposing the constraint that the neutrino and the
charged lepton four-momenta combine to form the $W$ rest
mass~\cite{STROUGHAIR,CHH}. Neglecting the lepton mass one finds
\begin{eqnarray}
\noalign{\vskip 5pt}
p_L^{}(\nu) &=& {1\over 2\, p^2_T(\ell) } \Biggl\{ p_L^{}(\ell)
\Bigl(M_W^2
+ 2\, \hbox{\bf p}_T^{}(\ell) \cdot \hbox{\bf p}\llap/_T^{} \Bigr)  \\
\noalign{\vskip 5pt}
&&\qquad\qquad \pm \, p(\ell)\, \biggl[ \Bigl(M_W^2
+ 2\, \hbox{\bf p}_T^{}(\ell) \cdot
\hbox{\bf p}\llap/_T^{} \Bigr)^2
- 4\, p_T^2(\ell)\, p\llap/_T^2 \biggr]^{1/2} \Biggr\}
\>,
\nonumber
\end{eqnarray}
where $p_L^{}(\ell)$ denotes the longitudinal momentum of the charged
lepton. The two solutions for $p_L^{}(\nu)$ are used to reconstruct
two values for $M_{WZ}^{}$. Both values are then
histogrammed, each with half the event weight.

The differential cross section for $M_{WZ}$ in the reaction
$p \, \bar p \rightarrow W^{+}Z + X \rightarrow \ell_1\nu_1
\ell_2^+\ell_2^-+X$ at $\sqrt{s} = 1.8$~TeV
is shown in Fig.~\ref{FIG:MWZTEV}.  The Born and NLO results are shown in
Fig.~\ref{FIG:MWZTEV}a and Fig.~\ref{FIG:MWZTEV}b, respectively.  In
both cases, results are displayed for the SM and for
three sets of anomalous couplings, namely,
$(\lambda^0 = -0.5,      \, \Delta\kappa^0 = \Delta g_1^0 = 0)$,
$(\Delta\kappa^0 = -1.0, \, \lambda^0 = \Delta g_1^0 = 0)$, and
$(\Delta g_1^0 = -0.5,   \, \Delta\kappa^0 = \lambda^0 = 0)$.
For simplicity, only one anomalous
coupling at a time is allowed to differ from its SM value.
The figure shows that at the Tevatron center of mass energy,
NLO QCD corrections do not have a large influence on the sensitivity of
the reconstructed $WZ$ invariant mass distribution to anomalous
couplings. The
${\cal O}(\alpha_s)$ corrections at Tevatron energies are approximately
30\% for the SM as well as for the anomalous coupling cases. Since the
anomalous terms in the helicity amplitudes grow like
$\sqrt{\hat s}/M_W$ ($\hat s/M_W^2$) for $\Delta\kappa$ ($\lambda$ and
$\Delta g_1$)~\cite{WILLEN}, where $\hat s$ denotes the parton center of
mass energy squared, non-standard couplings give large
enhancements in the cross section at large values of $M_{WZ}$.

The $WZ$ invariant mass distributions at the DiTevatron and the LHC are
shown in Figs.~\ref{FIG:MWZDITEV} and \ref{FIG:MWZLHC}, respectively.
In both cases, the sensitivity of the $M_{WZ}$ distribution to
anomalous $WWZ$ couplings is significantly more pronounced  than for
$p\bar p$ collisions at $\sqrt{s}=1.8$~TeV. NLO QCD corrections enhance
the SM $M_{WZ}$ differential cross section by about a factor~2 at the LHC,
whereas the ${\cal O}(\alpha_s)$ corrections at the DiTevatron are very
similar in size to those found at Tevatron energies. For non-standard
$WWZ$ couplings, the QCD corrections are more modest at the LHC.
Because of the form
factor parameters assumed, the result for $\Delta\kappa^0=-1$ approaches
the SM result at large values of $M_{WZ}$. As mentioned before, we have
used $n=2$ and a form factor scale of $\Lambda_{FF}=1$~TeV in all our
numerical simulations
[see Eqs.~(\ref{EQ:GONEFORM}) --~(\ref{EQ:LAMBDAFORM})].
For a larger scale $\Lambda_{FF}$, the
deviations from the SM result become more pronounced at high energies.
No significant change in the shape of the $M_{WZ}$ distribution is
observed for all center of mass energies considered.

In Fig.~\ref{FIG:MWZ} we investigate the influence of anomalous $WWZ$
couplings on the $WZ$ invariant mass spectrum at next-to-leading order,
together with the effect
of the twofold ambiguity in $M_{WZ}$ originating from the reconstruction
of the longitudinal momentum of the neutrino from the $W$ decay, in more
detail. The lower dotted, dashed, and dot-dashed lines display
the $WZ$ invariant mass distribution for $\Delta\kappa^0=+1$ (+1), $\Delta
g_1^0=+0.5$ (+0.25), and $\lambda^0=+0.5$ (+0.25), whereas the upper
curves show the $M_{WZ}$ spectrum for $\Delta\kappa^0=-1$ ($-1$),
$\Delta g_1^0=-0.5$ ($-0.25$), and $\lambda^0=-0.5$ ($-0.25$) at the
Tevatron (LHC). For $\Delta\kappa^0$ and $\Delta g_1^0$, negative anomalous
couplings lead to significantly larger deviations from the SM prediction
than positive non-standard couplings of equal magnitude, whereas there
is little difference for $\lambda^0$ (dashed lines). For
$\Delta\kappa^0$ the sign dependence is more pronounced at small
energies. Other distributions, such as
the cluster transverse mass distribution or the transverse momentum
distribution of the $Z$ boson display a similar behaviour.

This effect can be easily understood from the high energy behaviour of
the $WZ$ production amplitudes, ${\cal M}(\lambda_Z,\lambda_W)$, where
$\lambda_Z$ ($\lambda_W$) denotes the helicity of the $Z$ ($W$)
boson~\cite{WILLEN,UJH}. In the SM, only ${\cal M}(\pm,\mp)$ and ${\cal
M}(0,0)$ remain finite for $\hat s\to\infty$. Contributions to the
helicity amplitudes
proportional to $\lambda$ mostly influence the $(\pm,\pm)$ amplitudes
and increase like $\hat s/M_W^2$ at large energies. The SM $(\pm,\pm)$
amplitudes vanish like $1/\hat s$, and the non-standard terms
dominate except for the threshold region, $\sqrt{\hat s}\approx
M_W+M_Z$. For non-standard values of $\lambda$, the cross section
therefore depends only very little on the sign of the anomalous
coupling. Terms proportional to $\Delta g_1$ also increase like $\hat
s/M_W^2$ with energy, but mostly contribute to the (0,0) amplitude,
which remains finite in the SM in the high energy limit. Interference
effects between the SM and the anomalous contributions to the (0,0)
amplitude, thus, are non-negligible, resulting in a significant
dependence of the cross section on the sign of $\Delta g_1$. Terms
proportional to $\Delta\kappa$, finally, are proportional to $\sqrt{\hat
s}/M_W$ and mostly influence the amplitudes with a longitudinal $W$ and
a transverse $Z$ boson. In the SM, these terms vanish like $1/\sqrt{\hat
s}$. In the high energy limit one therefore expects little dependence of
the cross section on the sign of $\Delta\kappa$, similar to the
situation encountered for $\lambda$ (see Fig.~\ref{FIG:MWZ}b). However,
since the terms
proportional to $\Delta\kappa$ increase less drastically with energy,
interference effects between those terms and the SM amplitudes are
substantial near threshold.

The dash-double-dotted line in Fig.~\ref{FIG:MWZ} shows the true
$M_{WZ}$ distribution. The distribution of the reconstructed invariant
mass at both center of mass energies is harder than the true $M_{WZ}$
distribution. At LHC energies, the twofold ambiguity in the
reconstructed $WZ$ invariant mass only slightly affects the $M_{WZ}$
distribution (Fig.~\ref{FIG:MWZ}b). At the Tevatron, on the other hand,
the true and reconstructed invariant masses are quite different for
$M_{WZ}>500$~GeV, thus severely degrading the sensitivity to non-standard
$WWZ$ couplings. If the $W$ decay is treated in the narrow width
approximation, one of the two reconstructed invariant masses coincides
with the true $WZ$ invariant mass.
Since the $M_{WZ}$ spectrum is steeply falling,
the incorrect solution of the reconstructed invariant mass influences
the distribution in a noticeable way only if it is larger than the true $WZ$
invariant mass. The average difference (absolute value) between the two
reconstructed values of $M_{WZ}$ is almost independent of the center of
mass energy. As a result, the twofold ambiguity in the reconstructed
$WZ$ invariant mass affects the $M_{WZ}$ spectrum at the LHC much less
than at the Tevatron.

As an alternative to the $WZ$ invariant mass spectrum, the differential
cross section of the cluster transverse mass
$M_T(\ell_1\ell_2^+\ell_2^-;p\llap/_T)$~\cite{CTMASS} can be studied.
The cluster transverse mass is defined by
\begin{eqnarray}
M_T^2 (c;p\llap/_T^{}) & = &
\biggl[ \Bigl( M_{c}^2
 + \bigl| \hbox{\bf p}_{T}^{}(c) \bigr|^2 \Bigr)^{1/2}
 + p\llap/_T^{} \biggr]^2
- \bigl|   \hbox{\bf p}_{T}^{}(c)
         + \hbox{\bf p}\llap/_T^{} \bigr|^2 \, , \> \>
\end{eqnarray}
where $M_{c}$ denotes the invariant mass of the cluster
$c=\ell_1\ell_2^+\ell_2^-$. The $M_T$ distribution
at the Tevatron and the LHC is shown in Fig.~\ref{FIG:MTCL}. Since
QCD corrections change its shape only slightly, we only show the NLO
$M_T$ distribution. At the Tevatron, the cluster transverse mass
distribution is seen to be significantly more
sensitive to anomalous couplings than the reconstructed $WZ$ invariant
mass distribution, in particular for $\lambda$ and $\Delta g_1$.
The cluster transverse mass distribution for $p\bar p$
collisions at $\sqrt{s}=3.5$~TeV is qualitatively very similar to
that obtained at Tevatron energies and is therefore not shown.

In Figs.~\ref{FIG:PTTEV} --~\ref{FIG:PTLHC} we show the differential
cross section for the transverse momentum of the $Z$ boson, $p_T^{}(Z)$.
The $p_T^{}(Z)$ spectrum is seen
to be considerably more sensitive to non-standard $WWZ$ couplings than
the cluster transverse mass or the $WZ$ invariant mass distributions.
At high transverse momentum, a large enhancement of the cross section is
observed. On the other hand, at Tevatron and DiTevatron energies,
the $p_T^{}(Z)$ differential cross section is smaller than predicted in
the SM for $p_T^{}(Z)<30$~GeV, if anomalous $WWZ$ couplings are present.
Due to the relatively
large $p\llap/_T$ cut imposed, the $Z$ boson transverse momentum
distribution at the LHC in the Born approximation displays a pronounced
shoulder at $p_T^{}(Z)\approx
65$~GeV (see Fig.~\ref{FIG:PTLHC}a). The $p\llap/_T$ cut mostly
affects the small $p_T^{}(Z)$ region ($p_T^{}(Z)<100$~GeV) and
therefore does
not significantly reduce the sensitivity to non-standard $WWZ$ couplings.
Once NLO QCD corrections are taken
into account this shoulder disappears due to contributions from the
real emission subprocesses, $q_1\bar q_2\to WZg$ and $q_1g\to WZq_2$.

In contrast to the other distributions studied so far, the shape of the
SM $Z$ boson transverse momentum spectrum is considerably affected by NLO
QCD corrections. This is demonstrated in detail in Fig.~\ref{FIG:KFAC},
where we show the ratio of the NLO and LO differential cross sections of
the $Z$ boson transverse momentum. At Tevatron and DiTevatron energies, the
${\cal O}(\alpha_s)$ corrections are approximately 30\% for the SM, and
35~--~40\% for the anomalous coupling cases at small $p_T^{}(Z)$ values. In
the SM case, the size of the QCD corrections increases to $\sim 60\%$
for $p_T^{}(Z)=300$~GeV at the Tevatron, and to $\sim 80\%$ at the
DiTevatron. For non-standard couplings, on the other hand, the
QCD corrections are between 20\% and 40\% over the whole $p_T^{}(Z)$ range
plotted. This is exemplified by the dashed and dot-dashed lines in
Fig.~\ref{FIG:KFAC}a, which show the ratio of NLO to LO cross sections
for $\lambda^0=-0.5$ at the Tevatron and DiTevatron, respectively.
At the LHC (see Fig.~\ref{FIG:KFAC}b), the shape of the
$p_T^{}(Z)$ distribution is drastically altered by the NLO QCD corrections.
At $p_T^{}(Z)=800$~GeV, the QCD corrections increase the SM cross section by
about a factor~5, whereas the enhancement is only a factor~1.6 at
$p_T^{}(Z)=100$~GeV. In the presence of anomalous
couplings, the higher order QCD corrections are much smaller than in the
SM. In regions where the anomalous terms dominate, the ${\cal
O}(\alpha_s)$ corrections are typically between 20\% and 40\%. This is
illustrated by the dashed curve in Fig.~\ref{FIG:KFAC}b, which shows the
NLO to LO cross section ratio for $\lambda^0=-0.25$.
At next-to-leading order, the sensitivity of the $Z$ boson transverse
momentum spectrum to anomalous couplings is thus considerably reduced.
The transverse momentum distribution of the $W$ boson in the SM
exhibits a similar strong sensitivity to
QCD corrections. Qualitatively, the change of the shape of the $p_T^{}(Z)$
distribution in $WZ$ production and of the photon transverse momentum
distribution in $W\gamma$ production at high energies~\cite{BHO} are
very similar.

\subsection{Exclusive NLO QCD Corrections}

The large QCD corrections at high values of $p_T^{}(Z)$ are caused by the
combined effects of destructive interference in the Born process, a
collinear enhancement factor in the $q_1g\to WZq_2$ partonic cross
section for $p_T^{}(Z)\gg M_W$, and the large $qg$ luminosity at LHC
energies. In the SM, delicate cancellations between the amplitudes of
the Born diagrams occur in the central rapidity region in $WZ$
production. These cancellations are responsible for the approximate
amplitude zero~\cite{UJH} and suppress the
$WZ$ differential cross section, in particular for large $W$ and $Z$
transverse momenta. In the limit $p_T^{}(Z)\gg M_W$, the cross section for
$q_1g\rightarrow WZ q_2$ can be obtained using the Altarelli-Parisi
approximation for collinear emission. One finds~\cite{FRIX}:
\begin{eqnarray}
\noalign{\vskip 5pt}
d\hat\sigma(q_1g\rightarrow WZ q_2)=d\hat\sigma(q_1g\rightarrow
q_1Z)\,
{g^2_W\over 16\pi^2}\,\ln^2\!\left({p_T^2(Z)\over M_W^2}\right)\, ,
\label{EQ:COLLAPPROX}
\end{eqnarray}
where $g_W^{}=e/\sin\theta_{\rm W}$.
Thus, the quark gluon fusion process carries an enhancement factor
$\ln^2(p_T^2(Z)/M_W^2)$ at large values of the $Z$ boson transverse momentum.
It arises from the kinematic region where the $Z$ boson is
produced at large $p_T^{}$ and recoils against the quark, which radiates a
soft $W$ boson collinear to the quark. Since the Feynman
diagrams entering the derivation of Eq.~(\ref{EQ:COLLAPPROX})
do not involve the $WWZ$ vertex, the logarithmic enhancement
factor only affects the SM matrix elements. At the LHC, the
$p_T^{}(Z)$ differential cross section obtained
using Eq.~(\ref{EQ:COLLAPPROX}) agrees within 30\% with the exact $Z$
boson transverse momentum distribution for $p_T^{}(Z)
>200$~GeV~\cite{FRIX}.
Together with the very large $qg$ luminosity at supercollider
energies and the suppression of the SM $WZ$ rate at large values of the
$Z$ boson transverse momentum in the Born approximation, the logarithmic
enhancement factor is responsible for the size of the inclusive NLO QCD
corrections to $WZ$ production, as well as for the change in the
shape of the $p_T^{}(Z)$ distribution. The same enhancement factor
also appears in the antiquark gluon fusion process, however, the $\bar
qg$ luminosity is much smaller than the $qg$ luminosity for large
values of the $Z$ boson
transverse momentum. Since the $W$ boson does not couple directly to
the gluon, the process $q_1\bar q_2\to WZ g$ is not enhanced at large
values of the $Z$ boson
transverse momentum. Arguments similar to those presented above also
apply to the $W$ boson transverse momentum distribution in the limit
$p_T^{}(W) \gg M_Z$, and a relation analogous to Eq.~(\ref{EQ:COLLAPPROX})
can be derived.

{}From the picture outlined in the previous paragraph,
one expects that, at next-to-leading
order at supercollider energies, $WZ$ events with a high $p_T^{}$
$Z$ boson most of the time also contain a high transverse momentum jet. At
the Tevatron, on the other hand, the fraction of high $p_T^{}(Z)$
$WZ$ events with a hard jet should be considerably smaller, due to
the much smaller $qg$ luminosity at lower energies. The decomposition
of the inclusive SM NLO $p_T^{}(Z)$ differential cross section into NLO
0-jet and LO 1-jet exclusive cross sections at the Tevatron
and LHC is shown in Figs.~\ref{FIG:PTZTEVEX}a and~\ref{FIG:PTZLHCEX}a,
respectively. The SM NLO 0-jet $p_T^{}(Z)$ distribution at the two center
of mass energies are compared with
the corresponding distributions obtained in the Born approximation in
Figs.~\ref{FIG:PTZTEVEX}b and~\ref{FIG:PTZLHCEX}b.
Here, a jet is defined as a quark or gluon with
\begin{eqnarray}
p_T^{}(j)>10~{\rm GeV}\hskip 1.cm {\rm and} \hskip 1.cm |\eta(j)|<2.5
\label{EQ:TEVJET}
\end{eqnarray}
at the Tevatron and
\begin{eqnarray}
p_T^{}(j)>50~{\rm GeV}\hskip 1.cm {\rm and} \hskip 1.cm |\eta(j)|<3
\label{EQ:SSCJET}
\end{eqnarray}
at the LHC. The sum of the NLO 0-jet and the LO 1-jet exclusive cross
section is equal to the inclusive NLO cross section.
The results for the NLO exclusive $WZ+0$~jet and the LO exclusive
$WZ+1$~jet differential cross sections depend explicitly on the jet
definition. Only the inclusive NLO distributions are independent of the
jet definition. It should be noted that
the jet transverse momentum threshold can not be lowered to
arbitrarily small values in our calculation for theoretical reasons.
For transverse momenta below 5~GeV (20~GeV) at the Tevatron
(LHC), soft gluon resummation effects are expected to significantly change
the jet $p_T^{}$ distribution~\cite{RESUM}. For the jet definitions listed
above (Eqs.~(\ref{EQ:TEVJET}) and~(\ref{EQ:SSCJET})), these effects
are expected to be unimportant, and therefore are ignored in our
calculation.

With the jet definition of Eq.~(\ref{EQ:TEVJET}), the inclusive NLO
cross section at the Tevatron is composed predominately of 0-jet events
(see Fig.~\ref{FIG:PTZTEVEX}a). Due to the logarithmic enhancement
factor, the 1-jet cross section becomes relatively more important at
large values of the $Z$ boson transverse momentum. For $p_T^{}(Z)$ values
above
200~GeV, the 1-jet cross section is larger than the 0-jet rate at the
LHC, and dominates completely at high $p_T^{}(Z)$
(see Fig.~\ref{FIG:PTZLHCEX}a).
Figure~\ref{FIG:PTZTEVEX}b compares the NLO $WZ+0$~jet cross section
with the result obtained in the Born approximation at the Tevatron. For
the jet definition chosen [see Eq.~(\ref{EQ:TEVJET})], the results are
almost identical over the entire transverse momentum range displayed. At
the LHC, the NLO $WZ+0$~jet result is at most 20\%
smaller than the cross section obtained in the Born approximation
(Fig.~\ref{FIG:PTZLHCEX}b). Note that the
characteristic shoulder at $p_T^{}(Z)\approx 65$~GeV in the Born $p_T^{}(Z)$
distribution, which results from the large $p\llap/_T$ cut, is
eliminated in the NLO $WZ+0$~jet differential cross section. The LO and
NLO $WZ+0$~jet $p_T^{}(\ell_{1,2})$, $p\llap/_T$, and $y(\ell_{1,2})$
differential
cross sections also agree to better than 20\%~\cite{NEWJO}. The results
for DiTevatron energies are very similar to those obtained for the
Tevatron.

If the jet defining $p_T^{}$ threshold is lowered to 30~GeV
and jets can be identified out to $|\eta(j)|=4.5$ at the LHC, the NLO
$WZ+0$~jet $p_T^{}(Z)$ differential cross section is approximately 30\%
smaller than the result obtained with the jet definition of
Eq.~(\ref{EQ:SSCJET}).
Present studies suggest~\cite{ATLAS,CMS,LHCJET} that jets fulfilling the
criteria of Eq.~(\ref{EQ:SSCJET}) can be identified at the LHC without
problems, whereas it will be difficult to reconstruct a jet with a
transverse momentum of less than 30~GeV. The pseudorapidity range
covered by the LHC is not expected to extend beyond $|\eta|=4.5$.

The results shown so far were obtained for $Q^2=M^2_{WZ}$. Since the
$WZ+1$~jet and the $WZ+0$~jet cross section in the Born approximation
are tree level results, they are sensitive to the choice of the
factorization scale $Q^2$. Figure~\ref{FIG:QSCALE} displays
the scale dependence of the Born, the inclusive NLO, the ${\cal
O}(\alpha_s)$ 0-jet exclusive, and the 1-jet exclusive cross sections
for the Tevatron and LHC center of mass energies. The total cross
section for the reaction
$p\,p\hskip-7pt\hbox{$^{^{(\!-\!)}}$} \rightarrow W^+Z + X
\rightarrow \ell_1^+\nu_1\ell_2^+\ell_2^- + X$ is plotted versus the
scale $Q$. The factorization scale $M^2$ and the renormalization scale
$\mu^2$ have both been set equal to $Q^2$.

The scale dependence of the Born cross section enters only through the
$Q^2$ dependence of the parton distribution functions. The qualitative
differences between the results at the Tevatron and the LHC
are due to the differences between $p \bar p$ versus $pp$ scattering
and the ranges of the $x$-values probed.  At the Tevatron, $WZ$
production in $p \bar p$ collisions is dominated by valence quark
interactions.  The valence quark distributions decrease with $Q^2$
for the $x$-values probed at the Tevatron (typically $x > 0.1$).
On the other
hand, at the LHC, sea quark interactions dominate in the $pp$
process and smaller $x$-values are probed (typically $x \sim 0.02$).
The sea quark
distributions increase with $Q^2$ for the $x$-values probed at the
LHC.  Thus the Born cross section decreases with $Q^2$
at the Tevatron but increases with $Q^2$ at the LHC.

The scale dependence of the 1-jet exclusive cross section enters via
the parton distribution functions and the running coupling $\alpha_s(Q^2)$.
Note that the 1-jet exclusive cross section is calculated only to
lowest order and thus exhibits a strong scale dependence.
The dependence on $Q$ here is dominated by the scale dependence
of $\alpha_s(Q^2)$ which is a decreasing function of $Q^2$.
At the NLO level, the $Q$ dependence enters not only via the
parton distribution functions and the running coupling $\alpha_s(Q^2)$,
but also through explicit factorization scale dependence in the
order $\alpha_s(Q^2)$ correction terms.
The NLO 0-jet exclusive cross section is almost independent of the scale
$Q$. It shows a non-negligible variation with the scale only in the
region $Q<100$~GeV at the Tevatron. In the $WZ+0$~jet cross section,
the scale dependence of the
parton distribution functions is compensated by that of $\alpha_s(Q^2)$
and the explicit factorization scale dependence in the correction terms.
The $Q$ dependence of the inclusive NLO cross section is dominated by the
1-jet exclusive component and is significantly larger than that of the NLO
0-jet cross section.

\subsection{Sensitivity Limits}

We now study the impact that NLO QCD corrections to $WZ$ production
have on the
sensitivity limits for $\Delta g_1^0$,
$\Delta\kappa^0$, and $\lambda^0$ at the
Tevatron, DiTevatron, and LHC. For the Tevatron we
consider integrated luminosities of $\int\!{\cal L}dt=1$~fb$^{-1}$,
as envisioned for the
Main Injector era, and 10~fb$^{-1}$ (TeV*) which could be achieved
through additional upgrades of the Tevatron accelerator complex~\cite{GPJ}.
In the case of the DiTevatron we assume an integrated
luminosity of 10~fb$^{-1}$. For the LHC we consider integrated
luminosities of $\int\!{\cal L}dt=10$~fb$^{-1}$ and
100~fb$^{-1}$~\cite{LHC}.
To extract limits at the Tevatron, TeV*, and DiTevatron, we shall
sum over both $W$ charges. For the LHC, we only consider $W^+Z$
production. Interference effects between $\Delta g_1^0$,
$\Delta\kappa^0$, and $\lambda^0$ are fully incorporated in our analysis.

To derive 95\%~CL limits we use the $p_T^{}(Z)$ distribution and perform a
$\chi^2$ test~\cite{ZGAM}. In the Born approximation, the $Z$ boson
transverse momentum distribution in general yields the best sensitivity
bounds. Furthermore, we use the cuts summarized in Sec.~IIIB, and the
jet definitions of Eqs.~(\ref{EQ:TEVJET}) and~(\ref{EQ:SSCJET}). Unless
explicitly stated otherwise, a dipole form factor ($n=2$) with scale
$\Lambda_{FF}=1$~TeV is assumed.
For the Tevatron with 1~fb$^{-1}$ we split the $p_T^{}(Z)$
distribution into~3 bins, whereas~7 bins are used in all other
cases. In each bin the
Poisson statistics are approximated by a Gaussian distribution. In order
to achieve a sizable counting rate in each bin, all events with
$p_T^{}(Z)> 60$~GeV (120~GeV) at the Tevatron (TeV*) are collected in a
single bin. Similarly, all events with $p_T^{}(Z)> 180$~GeV (240~GeV
[480~GeV]) at the DiTevatron (LHC with 10~fb$^{-1}$ [100~fb$^{-1}$])
are accumulated into one bin.
This procedure guarantees that a high statistical significance cannot
arise from a single event at large transverse momentum, where the
SM predicts, say, only 0.01 events. In order to derive
realistic limits we allow for a normalization uncertainty of 50\% in the
SM cross section. Background contributions are expected to be small for
the cuts we impose, and are ignored in our derivation of sensitivity bounds.

Our results are summarized in Figs.~\ref{FIG:PPBARLIM}
and~\ref{FIG:PPLIM}, and Tables~I and~II. The cross section in
each bin is a bilinear function of the anomalous couplings $\Delta\kappa^0$,
$\lambda^0$, and $\Delta g_1^0$. Studying the correlations in the
$\Delta\kappa^0$ --~$\lambda^0$, the $\Delta\kappa^0$ --~$\Delta g_1^0$,
and the $\Delta g_1^0$ --~$\lambda^0$ planes is therefore sufficient
to fully include all interference effects between the various
$WWZ$ couplings. Figure~\ref{FIG:PPBARLIM} shows 95\% CL contours in the
three planes for the $p\bar p$ collider options obtained from the
inclusive NLO $p_T^{}(Z)$ distribution. Table~I displays the 95\%~CL
sensitivity limits,
including all correlations, at leading order and next-to-leading order
for the three $WWZ$ couplings for the process $p\bar p\rightarrow W^\pm
Z+X\rightarrow\ell_1^\pm\nu_1\ell_2^+\ell_2^-+X$. At Tevatron and
DiTevatron energies, the increase of the cross section at ${\cal
O}(\alpha_s)$ and the reduced sensitivity at large values of the
$Z$ boson transverse momentum balance each other, and
the limits obtained at LO and NLO are usually very similar. Our limits
fully reflect the strong sign dependence of the differential cross
sections observed for $\Delta\kappa^0$ and $\Delta g_1^0$ (see
Fig.~\ref{FIG:MWZ}).

With an integrated luminosity of 1~fb$^{-1}$ it will not be possible to
perform a very precise measurement of the $WWZ$
vertex in the $WZ\to\ell_1\nu_1\ell_2^+\ell_2^-$ channel at the
Tevatron. For integrated luminosities of less than a few~fb$^{-1}$,
the limits which can be achieved, however,
may be significantly improved by combining the bounds from $WZ\to\ell_1\nu_1
\ell_2^+\ell_2^-$ with the limits obtained from $p\bar p\to
WZ\to\ell_1\nu_1 jj$ and $p\bar p\to WZ \to jj \ell_2^+\ell_2^- $ at large
di-jet transverse momenta~\cite{DPF}. Currently, these channels are
used by the CDF
Collaboration to extract information on the structure of the $WWZ$
vertex~\cite{GLAS,TH}. Decay modes where the $W$ or $Z$
boson decays hadronically have a considerably larger branching ratio
than the $WZ\to\ell_1\nu_1\ell_2^+\ell_2^-$ channel and thus yield
higher rates. On the other hand, they are plagued by a substantial
$W/Z+2$~jet
QCD background, which, for large integrated luminosities ($\geq
10$~fb$^{-1}$), will eventually limit the sensitivity of the
semi-hadronic $WZ$ decay channels to anomalous $WWZ$ couplings.

The limits which can be achieved for $\Delta\kappa^0$ at the TeV* from
$WZ\to\ell_1\nu_1\ell_2^+\ell_2^-$ are about a
factor~1.8 better than those at the Tevatron with 1~fb$^{-1}$. The
bounds on $\Delta g_1^0$ and $\lambda^0$ improve by a factor~2 to~2.7.
Increasing the energy of the Tevatron to 3.5~TeV (DiTevatron) improves
the limits again significantly, in particular, the bound on
$\Delta\kappa^0$. Due to the rather strong interference effects
between the SM and the anomalous terms of the helicity amplitudes for
$\Delta g_1$ and $\Delta\kappa$, the contours sometimes deviate
substantially from the elliptical form naively expected. Furthermore,
significant correlations are observed, in particular, between
$\Delta\kappa^0$ and $\Delta g_1^0$ (see Fig.~\ref{FIG:PPBARLIM}b). The
limits obtained with a 0-jet requirement imposed are virtually identical to
those resulting from the inclusive NLO $p_T^{}(Z)$ distribution.

The 95\%~CL limit contours for the LHC are shown in
Fig.~\ref{FIG:PPLIM}. Table~II summarizes the LO and NLO sensitivity
bounds which can be achieved at the LHC.
At supercollider energies, the inclusive NLO QCD corrections
in the SM are very large and drastically change the shape of the SM
$p_T^{}(Z)$
distribution (see Fig.~\ref{FIG:PTLHC}). As a result, NLO QCD corrections
reduce the sensitivity to anomalous couplings by 20~--~40\%.
As the integrated luminosity increases, larger transverse momenta
become accessible. The difference between the LO and NLO sensitivity
bounds for 100~fb$^{-1}$ therefore is somewhat larger than for 10~fb$^{-1}$.
For the parameters chosen, the inclusive NLO bounds which can
be obtained from $pp\rightarrow W^+Z+X\rightarrow
\ell_1^+\nu_1\ell_2^+\ell_2^-+X$ at $\sqrt{s}=14$~TeV with $\int\!{\cal
L}dt=10$~fb$^{-1}$ are quite similar to those which are expected from
the DiTevatron for $W^\pm Z$ production and equal integrated luminosity.

As we have seen in Sec.~IIIE, the size of the ${\cal O}(\alpha_s)$
QCD corrections at the LHC can be significantly reduced by vetoing
hard jets in the central rapidity region, {\it i.e.}, by imposing a
``zero jet'' requirement and considering only the $WZ+0$~jet channel.
A zero jet cut for example has been imposed in the CDF measurement
of the ratio of $W$ to $Z$ cross sections~\cite{RATIO} and the $W$ mass
measurement~\cite{WMASS}. The sensitivity limits obtained for the
$WZ+0$~jet channel at NLO are 10~--~30\% better than those obtained
in the inclusive NLO case and are quite often close to those
obtained from the leading order $p_T^{}(Z)$ distribution (see Table~II and
the dotted contours in Fig.~\ref{FIG:PPLIM}). The NLO
$WZ+0$~jet differential cross section is also more stable to variations
of the factorization scale $Q^2$ than the Born and inclusive NLO
$WZ+X$ cross sections (see Fig.~\ref{FIG:QSCALE}).
The systematic errors which originate from the choice of $Q^2$ will thus
be smaller for bounds derived from the NLO $WZ+0$~jet
differential cross section than those obtained from the inclusive
NLO $WZ+X$ or the Born cross section. The limits extracted from the
$WZ+0$~jet exclusive channel depend only negligibly on the jet
definition used.

The bounds which can be achieved
at the LHC improve by up to a factor~3 if an integrated luminosity of
100~fb$^{-1}$ can be achieved (dot-dashed contours in
Fig.~\ref{FIG:PPLIM}). Note that the $\Delta\kappa^0$ and $\Delta g_1^0$
limits are strongly correlated in this case. This effect is due to
the relatively small form
factor scale chosen ($\Lambda_{FF}=1$~TeV), which significantly suppresses
the non-standard terms in the helicity amplitudes at high energies.

At Tevatron (DiTevatron) energies, the
sensitivities achievable are insensitive to the exact form and
scale of the form factor for $\Lambda_{FF}>400$~GeV
($\Lambda_{FF}>800$~GeV). At the LHC, the situation is somewhat
different and the sensitivity bounds depend on the value chosen for
$\Lambda_{FF}$~\cite{WILLEN}. This is illustrated in
Table~IIc, where we list the bounds which can be achieved at the LHC
with $\int\!{\cal L}dt=100$~fb$^{-1}$ and a form factor scale of
$\Lambda_{FF}=3$~TeV. The limits for the higher scale are a factor~1.8 to~5
better than those found for $\Lambda_{FF}=1$~TeV with the same
integrated luminosity. For $\Lambda_{FF}>3$~TeV, the
sensitivity bounds depend only marginally on the form factor scale, due
to the very rapidly falling cross section at the LHC for parton center
of mass energies in the multi-TeV region. The dependence of the limits
on the cutoff scale $\Lambda_{FF}$ in the form
factor can be understood easily from Fig.~\ref{FIG:PTLHC}. The improvement in
sensitivity with increasing $\Lambda_{FF}$ is due to the additional events
at large $p_T^{}(Z)$ which are suppressed by the
form factor if the scale $\Lambda_{FF}$ has a smaller value.

To a lesser degree, the bounds also depend on the power $n$ in the form factor,
which we have assumed to be $n=2$. For example, the less drastic cutoff
for $n=1$ instead of $n=2$ in the form
factor allows for additional high $p_T^{}(Z)$ events and therefore leads
to a slightly increased sensitivity to the low energy values $\Delta\kappa^0$,
$\Delta g_1^0$, and $\lambda^0$. The sensitivity bounds listed in
Tables~I and~II can thus be taken as representative for a wide class of
form factors, including the case where constant anomalous couplings are
assumed for $M_{WZ}<\Lambda_{FF}$, but invariant masses above
$\Lambda_{FF}$ are ignored in deriving the sensitivity
bounds~\cite{FLS}.

The bounds derived in this section are quite conservative.
At the LHC, the limits can easily be improved by 10 -- 20\% if $W^-Z+X$
production is included.
Further improvements may result from using more powerful statistical
tools than the simple $\chi^2$ test we performed~\cite{GREG}.

\section{AMPLITUDE ZEROS, RAPIDITY CORRELATIONS, AND CROSS SECTION RATIOS}

Recently, it has been shown that the SM amplitude for $q_1\bar q_2
\rightarrow W^\pm Z$ at the Born level exhibits an approximate zero
at high energies, $\hat s\gg M_Z^2$, located at~\cite{UJH}
\begin{eqnarray}
\cos\Theta^* = \cos\Theta^*_0 \approx \pm {1\over 3}\tan^2\theta_{\rm W}
\approx \pm 0.1,
\label{EQ:ZERO}
\end{eqnarray}
where $\Theta^*$ is the scattering
angle of the $Z$ boson relative to the quark direction in the $WZ$
center of mass frame. The approximate zero is the combined result of
an exact zero in the dominant helicity amplitudes ${\cal M}(\pm,\mp)$, and
strong gauge cancellations in the remaining amplitudes. At high
energies, only the $(\pm,\mp)$ and $(0,0)$ amplitudes remain non-zero
in the SM.  The existence of the zero in ${\cal M}(\pm,\mp)$ at
$\cos\Theta^*\approx\pm 0.1$ is
a direct consequence of the contributing $u$- and $t$-channel fermion
exchange diagrams and the left-handed coupling of the $W$ boson to fermions.
Unlike the $W^\pm \gamma$ case with its massless photon kinematics, the
zero has an energy dependence which is, however, rather weak for
energies sufficiently above the $WZ$ mass threshold.

In this Section, we consider possible observable consequences of the
approximate zero in $WZ$ production in hadronic collisions and the
impact of NLO QCD corrections on the relevant quantities. All numerical
simulations are carried out using the parameters and cuts summarized in
Secs.~IIIA and~IIIB. For the form factor, we again assume a dipole form
factor ($n=2$) with scale $\Lambda_{FF}=1$~TeV (see
Eqs.~(\ref{EQ:GONEFORM}) --~(\ref{EQ:LAMBDAFORM})).
Since the approximate amplitude zero in $WZ$ production is similar in
nature to the well-known radiation zero in $W\gamma$ production,
analogous strategies can be applied to search for observable signals.
The radiation zero in $W\gamma$ production leads to a pronounced dip in
the rapidity distribution of the photon in the parton center of mass
frame, $d\sigma/dy^*(\gamma)$~\cite{BAURZEP}. The approximate zero in
the $WZ$ amplitude is therefore expected to manifest itself as a
dip in the corresponding $y^*(Z)$ distribution. Here,
\begin{eqnarray}
y^*(Z)={1\over 2}\,
\ln\left( {{1+\beta_Z\cos\theta^*} \over {1-\beta_Z\cos\theta^*}} \right) \>,
\label{EQ:YSTZ}
\end{eqnarray}
and
\begin{eqnarray}
\beta_Z=\left[1-{4M_Z^2\hat s\over (\hat s-M_W^2+M_Z^2)^2}\right ]^{1/2}
\end{eqnarray}
where, to lowest order, $\hat s=M^2_{WZ}$ is the squared parton center
of mass energy, and $\theta^*$ the scattering angle of the $Z$ boson with
respect to the beam direction in the parton center of mass rest frame.
For $pp$ collisions the dip is
centered at $y^*(Z)=0$. In $p\bar p$ collisions, the location of the
minimum is determined by $\cos\Theta^*_0$ of Eq.~(\ref{EQ:ZERO}) ,
the average $WZ$
invariant mass, and the fraction of events originating from sea quark
collisions. As can be seen from Figs.~\ref{FIG:MWZTEV}
and~\ref{FIG:MWZDITEV}, most of the cross section originates from the
region $\sqrt{\hat s}=200$ --~250~GeV. Valence quark collisions dominate
at both, Tevatron and DiTevatron energies. The minimum of the $y^*(Z)$
distribution is therefore expected to occur at $y^*(Z)\approx \pm 0.06$.

The $|y^*(Z)|$ distribution at the Tevatron and the LHC in Born
approximation is shown in Fig.~\ref{FIG:YSTAR}. The rapidity
distribution of the $Z$ boson in the parton center of mass frame at the
DiTevatron is qualitatively very similar to that found at Tevatron
energies and is therefore not shown. The SM $|y^*(Z)|$
distribution in the true $WZ$ rest frame (dash-double-dotted curves)
displays a pronounced dip at $|y^*(Z)|=0$, which originates from the
approximate amplitude zero. At Tevatron energies, the dip is quite
significant. However, since the unobservable longitudinal neutrino
momentum can only be determined with a twofold ambiguity and, on an
event to event basis, one does not know which solution is the correct
one, both solutions have to be considered for each event. Assigning half
of the event weight to each solution, the dip in the $|y^*(Z)|$
distribution using the
reconstructed $WZ$ rest frame is considerably filled in (solid lines).
NLO QCD corrections further diminish the significance of the dip.

In Fig.~\ref{FIG:YSTAR} we also display the effect of non-standard $WWZ$
couplings on the $|y^*(Z)|$ distribution (using the reconstructed $WZ$
rest frame). The lower dotted, dashed, and dot-dashed lines display
the $|y^*(Z)|$ distribution for $\Delta\kappa^0=+1$ (+1), $\Delta
g_1^0=+0.5$ (+0.25), and $\lambda^0=+0.5$ (+0.25), whereas the upper
curves show the $|y^*(Z)|$ spectrum for $\Delta\kappa^0=-1$ ($-1$),
$\Delta g_1^0=-0.5$ ($-0.25$), and $\lambda^0=-0.5$ ($-0.25$) at the
Tevatron (LHC). Non-standard $WWZ$ couplings eliminate the
approximate amplitude zero~\cite{UJH} and, in general, tend to fill in
the dip. However, due to the relatively strong interference between
standard and anomalous contributions to the helicity amplitudes for
$\Delta\kappa^0$ and $\Delta g_1^0$ at low energies, the dip may even
become more pronounced for certain (positive) values of these two
couplings at the Tevatron (see the lower dotted line in
Fig.~\ref{FIG:YSTAR}a).

{}From Fig.~\ref{FIG:YSTAR} it is obvious that the dip signaling the
approximate amplitude zero in $q_1\bar q_2\to WZ$ will be difficult to
observe in the $|y^*(Z)|$ distribution.
In $W\gamma$ production, correlations between the rapidities of the
photon and the charged lepton originating from the $W$ decay offer
better access to the SM radiation zero than the $y^*(\gamma)$
distribution~\cite{BEL}. Knowledge of the neutrino
longitudinal momentum, $p_L^{}(\nu)$, is not required in determining
these correlations, and thus
event reconstruction problems originating from the two possible
solutions for $p_L^{}(\nu)$ are automatically avoided.
In $2\to 2$ reactions rapidity differences are invariant
under boosts, $\Delta y(\gamma,W)=y(\gamma)-y(W)=y^*(\gamma)-y^*(W)$.
One therefore expects the rapidity difference distribution, $d\sigma/d\Delta
y(\gamma,W)$, to exhibit a dip signaling the SM radiation zero. In
$W^\pm\gamma$ production, the dominant $W$ helicity in the SM is
$\lambda_W = \pm 1$~\cite{STR}, implying that
the charged lepton from $W\to\ell\nu$ tends to be emitted
in the direction of the parent
$W$, and thus reflects most of its kinematic properties. As a result,
the dip signaling the SM radiation zero manifests itself also in the
$\Delta y(\gamma,\ell)=y(\gamma)-y(\ell)$ distribution.

The corresponding $y(Z)-y(\ell_1)$ distribution for $W^+Z$ production in
the Born approximation is shown in Fig.~\ref{FIG:RAPDIFF} (solid line).
Analogous to the situation encountered in $q_1\bar q_2\to W\gamma$, the
approximate zero in the $WZ$ amplitude leads to a dip in the $y(Z)-y(W)$
distribution~\cite{FRIX}, which is located at $y(Z)-y(W)\approx\pm
0.12$ ($=0$) for
$W^\pm Z$ production in $p\bar p$ ($pp$) collisions. However, in contrast
to $W\gamma$ production, none of the $W$ helicities dominates
in $WZ$ production\cite{STR}. The charged lepton, $\ell_1$, originating
from the $W$ decay, $W\to\ell_1\nu_1$, thus only partly reflects the
kinematical
properties of the parent $W$ boson. As a result, a significant part of
the correlation present in the $y(Z)-y(W)$ spectrum is lost, and only
a slight dip survives in the SM $y(Z)-y(\ell_1)$ distribution.
Due to the non-zero average rapidity difference between the lepton
$\ell_1$ and the parent $W$ boson, the location of
the minimum of the $y(Z)-y(\ell_1)$ distribution in $p\bar p$
collisions is slightly shifted to $y(Z)-y(\ell_1)\approx 0.5$. The
$y(Z)-y(\ell_1)$ distribution at the DiTevatron is qualitatively very
similar to that obtained for $p\bar p$ collisions at $\sqrt{s}=1.8$~TeV
(see Fig.~\ref{FIG:DELYPPBAR}).

The significance of the dip in the $y(Z)-y(\ell_1)$ distribution depends
to some extent on the cut imposed on $p_T^{}(\ell_1)$ and the missing
transverse momentum. Increasing (decreasing) the cut on $p_T^{}(\ell_1)$
($p\llap/_T$) tends to increase the probability that $\ell_1$
is emitted in the flight direction of the
$W$ boson, and thus enhances the significance of the dip. If the
$p\llap/_T>50$~GeV cut at the LHC could be reduced to 20~GeV, the dip
signaling the approximate zero in the $WZ$ production amplitude would be
strengthened considerably.

Although the $Z$ boson rapidity
can readily be reconstructed from the four momenta of the
lepton pair $\ell_2^+\ell_2^-$ originating from the $Z$ decay, it would
be experimentally easier to directly study the rapidity correlations
between the charged leptons originating from the $Z\to\ell_2^+\ell_2^-$ and
$W\to\ell_1\nu_1$ decays. The dotted lines in Fig.~\ref{FIG:RAPDIFF}
show the SM $y(\ell_2^-)-y(\ell_1^+)$ distribution for $W^+Z$
production in the Born approximation. Because none of the $Z$ or $W$
helicities dominates\cite{STR} in $q_1\bar q_2\to WZ$, the
rapidities of the leptons from $W$ and $Z$ decays are almost completely
uncorrelated, and essentially no trace of the dip signaling the approximate
amplitude zero is left in the $y(\ell_2^-)-y(\ell_1^+)$ distribution.
The $y(\ell_2^+)-y(\ell_1^+)$ spectrum almost coincides with the
$y(\ell_2^-)-y(\ell_1^+)$ distribution and is therefore not shown.

In Figs.~\ref{FIG:DELYPPBAR} and~\ref{FIG:DELYPP} we show the influence of
NLO QCD corrections and non-standard $WWZ$ couplings (at NLO) on the
$\Delta y(Z,\ell_1)=y(Z)-y(\ell_1)$ spectrum. At Tevatron energies, the
shape of the distribution is seen to be hardly influenced by the ${\cal
O}(\alpha_S)$ QCD corrections. At the DiTevatron, the significance of
the dip is slightly reduced. At
LHC energies, the dip is completely eliminated by the inclusive
QCD corrections. The NLO $WZ+0$~jet $\Delta y(Z,\ell_1)$ distribution,
however, is very similar to the leading order rapidity difference
distribution (see Fig.~\ref{FIG:DELYPP}a).

The effect of anomalous $WWZ$ couplings on the NLO $\Delta y(Z,\ell_1)$
distribution is exemplified by the dashed and dotted lines in
Fig.~\ref{FIG:DELYPPBAR} and in Fig.~\ref{FIG:DELYPP}b. Similar to the
situation encountered in the $|y^*(Z)|$ distribution, the dip in the
$\Delta y(Z,\ell_1)$ distribution at Tevatron and DiTevatron energies
may be more pronounced than in the SM for certain (positive) values of
$\Delta\kappa^0$. The shape of the $\Delta y(Z,\ell_1)$ distribution is
seen to be quite sensitive to the sign of $\Delta\kappa^0$
(Fig.~\ref{FIG:DELYPPBAR}). The same behaviour is observed
for $\Delta g_1^0$, whereas positive and negative values of $\lambda^0$
lead to a very similar $\Delta y(Z,\ell_1)$ distribution. In general,
non-standard $WWZ$ couplings tend to fill in the dip. In order not to
overburden the figures, curves for $\lambda^0$ and $\Delta g_1^0$ are
not shown in
Fig.~\ref{FIG:DELYPPBAR}. If deviations from the SM prediction were to be
observed, it would be difficult to determine the sign of an anomalous
coupling from the shape of the $WZ$ invariant mass
distribution, the cluster transverse mass spectrum, or the $p_T^{}(Z)$
distribution. For $\Delta\kappa^0$ and $\Delta g_1^0$, the pronounced
difference in shape of the $\Delta y(Z,\ell_1)$ distribution for
positive and negative values may aid in determining the
sign. The influence of non-standard $WWZ$ couplings on the exclusive NLO
$WZ+0$~jet distribution is shown in Fig.~\ref{FIG:DELYPP}b. Curves are
only shown for positive values of the anomalous couplings.

The error bars associated with the solid curves in
Figs.~\ref{FIG:DELYPPBAR} and~\ref{FIG:DELYPP}a indicate the expected
statistical uncertainties for an integrated luminosity of 10~fb$^{-1}$
at the Tevatron and DiTevatron, and for $\int\!{\cal L}dt=100$~fb$^{-1}$
at the LHC. It appears that the
approximate zero in the $WZ$ amplitude will be rather difficult to
observe in the $\Delta y(Z,\ell_1)$ distribution. However, if the LHC
is operated below its design luminosity of
${\cal L}=10^{34}$~cm$^{-2}$~s$^{-1}$, it may be possible to reduce the
$p\llap/_T$ cut. As mentioned above, the significance of the dip in
the $\Delta y(Z,\ell_1)$ distribution increases if the missing
transverse momentum cut is lowered. The smaller collision rate is at
least partially compensated by the larger total cross section for the
reduced $p\llap/_T$ cut. It is thus possible that
the conditions to detect the dip in the $\Delta y(Z,\ell_1)$ distribution
improve if the LHC is operated below its design luminosity.
However, more detailed simulations are required before definite
conclusions can be drawn.

As an alternative to rapidity correlations, cross section ratios can be
studied. Many experimental
uncertainties, for example those associated with the lepton detection
efficiencies, or the uncertainty in the integrated luminosity are
expected to cancel, at least partially, in cross section ratios.
In Ref.~\cite{BEO}, the ratio of the $Z\gamma$ to $W\gamma$ cross sections,
${\cal R}_{Z\gamma/W\gamma}$, was shown to reflect
the radiation zero present in the SM $W\gamma$ helicity amplitudes. Due
to the radiation zero, the $W\gamma$ cross section is reduced in the
central rapidity region. With increasing photon transverse momenta,
events become more and more central in rapidity. The reduction of the
$W\gamma$ cross
section at small rapidities originating from the radiation zero thus
becomes more pronounced at high $p_T^{}(\gamma)$. This causes the photon
transverse momentum distribution of $q_1\bar q_2\to W\gamma$ to fall
significantly faster than the $p_T^{}(\gamma)$ spectrum of $q\bar q\to
Z\gamma$ where no radiation zero is present. As a result,
${\cal R}_{Z\gamma/W\gamma}$
increases rapidly with the minimum transverse momentum of the photon.

In Fig.~\ref{FIG:RZZ} we study the cross section ratio
\begin{eqnarray}
\noalign{\vskip 5pt}
{\cal R}_{ZZ/WZ}={B^2(Z\to\ell^+\ell^-)\,\sigma(ZZ)\over
B(Z\to\ell^+\ell^-)\,B(W\to\ell\nu)\,\sigma(W^\pm Z)}=
{B(Z\to\ell^+\ell^-)\,\sigma(ZZ)\over B(W\to\ell\nu)\,\sigma(W^\pm Z)}~,
\label{EQ:RATZZ}
\end{eqnarray}
as a function of the minimum transverse momentum of the $Z$ boson,
$p_T^{\rm min}$. To calculate the $ZZ$ cross section, we use the
results of Ref.~\cite{NEWJO} and assume the SM to be valid. The
$ZZ$ helicity amplitudes do not exhibit any zeros, whereas the SM $WZ$
amplitude shows an approximate zero in the central rapidity region. The
situation is thus qualitatively very similar to that encountered in the
ratio of $Z\gamma$ to $W\gamma$ cross sections, and one expects ${\cal
R}_{ZZ/WZ}$ to grow with $p_T^{\rm min}$. Figure~\ref{FIG:RZZ}a
demonstrates that, at Tevatron energies, ${\cal R}_{ZZ/WZ}$ indeed
rises quickly for $p_T^{\rm min}>100$~GeV in the SM, indicating the
presence of the approximate zero in the $WZ$ amplitude. For smaller
values of the minimum $Z$ boson
transverse momentum, ${\cal R}_{ZZ/WZ}$ is approximately constant.
In the low $p_T^{\rm min}$ region, the shape of the $p_T^{}(Z)$
distribution is dominated by $Z$ mass effects which are similar in
both processes. The cross section ratio
at next-to-leading order differs only by about 10\% from the LO ratio.

At LHC energies, the situation is more complex. For $p_T^{\rm
min}<100$~GeV, ${\cal R}_{ZZ/WZ}$ drops sharply due to the large
$p\llap/_T$ cut imposed, which significantly suppresses the $WZ$ cross
section. While the cross section ratio slowly rises with $p_T^{\rm
min}$ for $p_T^{\rm min}>100$~GeV at
leading order, ${\cal R}_{ZZ/WZ}$ continues to decrease if
inclusive NLO QCD corrections are taken into account
(Fig.~\ref{FIG:RZZ}b). The relatively slower rise of ${\cal R}_{ZZ/WZ}$
at LO at the LHC is due to the larger fraction of the cross sections
originating from sea quark collisions, and the different $x$-ranges
probed at the Tevatron and LHC. For $p_T^{\rm
min}(Z)=1$~TeV, the inclusive NLO cross section ratio is about a factor~3
smaller than ${\cal R}_{ZZ/WZ}$ at leading order. At large
values of the $Z$
boson transverse momentum, the QCD corrections to $WZ$ production at LHC
energies are substantially larger than in the $ZZ$ case~\cite{NEWJO},
resulting in a large discrepancy between the LO and NLO prediction for
${\cal R}_{ZZ/WZ}$. In contrast to the situation encountered at the
Tevatron, higher order QCD corrections completely blur the signal of the
approximate amplitude zero in the $WZ$ channel. Their size, however, can be
substantially reduced by imposing a zero jet
requirement (see Sec.~IIIE and Ref.~\cite{NEWJO}). The result for the
$ZZ+0$~jet to $W^\pm Z+0$~jet cross section ratio at the LHC is given by
the dotted line in Fig.~\ref{FIG:RZZ}b.  With a jet veto imposed, the NLO
$ZZ$ to $WZ$ cross section ratio rises with the minimum $Z$ boson transverse
momentum for $p_T^{\rm min}>100$~GeV, and differs by at most 15\% from
the LO prediction. At the Tevatron, the NLO 0-jet cross section ratio
virtually coincides with the ratio obtained at LO.

The dot-dashed curve in Fig.~\ref{FIG:RZZ}, finally, shows the $ZZ$ to
$WZ$ cross section ratio for $\Delta\kappa^0=+1$, illustrating the
behaviour of ${\cal R}_{ZZ/WZ}$ in presence of anomalous $WWZ$
couplings. At the Tevatron
(Fig.~\ref{FIG:RZZ}a), the dot-dashed curve has been calculated taking
into account inclusive ${\cal O}(\alpha_s)$ QCD corrections. At
LHC energies (Fig.~\ref{FIG:RZZ}b) the NLO $ZZ$ to $WZ$ cross section
ratio is plotted with a jet veto included.  Non-standard couplings
lead to an enhancement of the $WZ$ cross section, in particular at large
values of $p_T^{}(Z)$ and, at the Tevatron, ${\cal R}_{ZZ/WZ}$
decreases with $p_T^{\rm min}$. Due to the form factor parameters
assumed ($n=2$ and $\Lambda_{FF}=1$~TeV), the cross section ratio at the
LHC displays a broad minimum at $p_T^{\rm min}(Z)\approx 300$~GeV, and
increases quickly at large values of $p_T^{\rm min}$. For larger values of
$\Lambda_{FF}$, and/or non-zero values of $\Delta g_1^0$ or $\lambda^0$,
${\cal R}_{ZZ/WZ}$ rises more slowly, or may even decrease with
$p_T^{\rm min}(Z)$. In general, the $ZZ$ to $WZ$ cross section ratio as a
function of the minimum $Z$ boson transverse momentum differs
substantially in shape from the SM prediction for ${\cal R}_{ZZ/WZ}$
in presence of non-standard $WWZ$ couplings.

At the Tevatron, the limited number of $ZZ$ and $WZ$ events expected
in the purely leptonic channels will unfortunately limit the
usefulness of ${\cal R}_{ZZ/WZ}$. Even for an integrated
luminosity of 10~fb$^{-1}$ only a handful of events are expected for
$p_T^{}(Z)>150$~GeV, and it will be very difficult to establish the growth
with $p_T^{\rm min}(Z)$ predicted by the SM. At the LHC, the statistical
errors are expected to be much smaller, however, one can only hope to
observe the rise of ${\cal R}_{ZZ/WZ}$ signalling the
presence of the approximate zero in the $WZ$ channel if a 0-jet
requirement is imposed. Moreover, the rise of the cross section ratio is
very slow, and for $p_T^{\rm min}=600$~GeV only about~5 (2)~purely leptonic
$W^\pm Z$ ($ZZ$) events are expected for $\int\!{\cal
L}dt=100$~fb$^{-1}$. Combined, these effects will
make it quite difficult to accurately determine the slope of ${\cal
R}_{ZZ/WZ}$.

A cross section ratio which
suffers somewhat less from the small number of events expected at the
Tevatron, and which is less sensitive to QCD corrections at the LHC is
the ratio of $WZ$ to $W\gamma$ cross sections,
\begin{eqnarray}
\noalign{\vskip 5pt}
{\cal R}_{WZ/W\gamma}={B(Z\to\ell^+\ell^-)\,B(W\to\ell\nu)\,
\sigma(W^\pm Z)\over B(W\to\ell\nu)\,\sigma(W^\pm\gamma)}=
B(Z\to\ell^+\ell^-)~{\sigma(W^\pm Z)\over\sigma(W^\pm\gamma)}
\label{EQ:RATWG}
\end{eqnarray}
considered as a function of the minimum transverse momentum of the $Z$
boson and photon, $p_T^{\rm min}$, respectively. ${\cal R}_{WZ/W\gamma}$
measures the relative strength of the approximate zero in $q_1\bar
q_2\to WZ$ and the radiation zero in $W\gamma$ production.
Figure~\ref{FIG:RWG} shows the ratio ${\cal R}_{WZ/W\gamma}$ as a function
of $p_T^{\rm min}$ for the Tevatron (part a) and LHC (part b) center of
mass energies. In obtaining ${\cal R}_{WZ/W\gamma}$, we have considered
both the electron and muon decay channels of the $W$ and $Z$ bosons.
The $W\gamma$ cross section in Fig.~\ref{FIG:RWG} has been
calculated using the results of Ref.~\cite{BHO} and the following cuts
on the photon:
\begin{eqnarray}
p_T^{}(\gamma)> 10~{\rm GeV}, & \qquad\qquad & |\eta(\gamma)|<1,~~({\rm
Tevatron})  \nonumber\\
p_T^{}(\gamma)> 100~{\rm GeV}, & \qquad\qquad & |\eta(\gamma)|<2.5,~~({\rm
LHC})  \label{EQ:WGCUT}\\
M_T(\ell\gamma;p\llap/_T)>90~{\rm GeV},  &  \qquad\qquad &
\Delta R(\gamma,\ell) > 0.7. \nonumber
\end{eqnarray}

At small values of $p_T^{\rm min}$, the $WZ$ to $W\gamma$ cross section
ratio rises very rapidly, due to the finite $Z$ mass effects which dominate
the shape of the $p_T^{}(Z)$ spectrum in this region for $WZ$ production.
For $p_T^{\rm min}>100$~GeV (200~GeV) at the Tevatron (LHC),
${\cal R}_{WZ/W\gamma}$ is almost constant and independent of the center
of mass energy, indicating that the radiation zero in $q_1\bar q_2\to
W\gamma$ and the approximate
amplitude zero in $WZ$ production affect the respective photon and $Z$ boson
transverse momentum
distribution in a very similar way. At the Tevatron, NLO QCD corrections
reduce ${\cal R}_{WZ/W\gamma}$ by about 10\%. At LHC energies, the
individual ${\cal O}(\alpha_s)$ QCD corrections are very large for both
$WZ$ and $W\gamma$ production~\cite{BHO}, in particular at high transverse
momenta (see Figs.~\ref{FIG:PTLHC} and~\ref{FIG:PTZLHCEX}). In the
cross section ratio, these large corrections cancel almost
completely. For $p_T^{\rm min}>200$~GeV, QCD corrections reduce
${\cal R}_{WZ/W\gamma}$ by 20\% or less. In contrast to the LO cross
section ratio, which is completely flat for $p_T^{\rm
min}>200$~GeV, ${\cal R}_{WZ/W\gamma}$ at NLO slowly rises with
$p_T^{\rm min}$ at the LHC.

Most theoretical models with non-standard $WWZ$ couplings also
predict anomalous $WW\gamma$ couplings at the same time (see {\it
e.g.}, Ref.~\cite{HISZ}). The effects of
anomalous $WWZ$ and $WW\gamma$ couplings may cancel
almost completely in ${\cal R}_{WZ/W\gamma}$ if the $WWZ$ and $WW\gamma$
couplings are similar in magnitude and originate from operators of the
same dimension. This is illustrated by the dot-dashed and
dotted lines in Fig.~\ref{FIG:RWG}, which show ${\cal R}_{WZ/W\gamma}$
at LO and NLO for $\Delta\kappa_\gamma^0=\Delta\kappa^0=-1$. Here the
anomalous $WW\gamma$ coupling $\Delta\kappa_\gamma$ is defined through an
effective Lagrangian analogous to that of Eq.~(\ref{EQ:LAGRANGE}), and
we assume equal form factor scales and powers ($\Lambda_{FF}=1$~TeV and
$n=2$) for $\Delta\kappa$ and $\Delta\kappa_\gamma$. Both couplings
correspond to operators of dimension four in the effective Lagrangian.
Although the individual $p_T^{}(\gamma)$ and $p_T^{}(Z)$ differential cross
sections are enhanced by up to one order of magnitude (see {\it e.g.},
Figs.~\ref{FIG:PTTEV} and~\ref{FIG:PTLHC}), ${\cal R}_{WZ/W\gamma}$
agrees to better than 20\% with the NLO SM cross section ratio for
$\Delta\kappa_\gamma^0=\Delta\kappa^0=-1$.

At DiTevatron energies, the results for ${\cal R}_{ZZ/WZ}$ and ${\cal
R}_{WZ/W\gamma}$ are qualitatively similar to those obtained for
Tevatron and are therefore not shown.

\section{SUMMARY}

$WZ$ production in hadronic collisions provides an opportunity to
probe the structure of the $WWZ$ vertex in a direct and
essentially model
independent way. Previous studies of $p\,p\hskip-7pt\hbox{$^{^{(\!-\!)
}}$} \rightarrow W^{\pm}Z$~\cite{WILLEN,WWZ} have been
based on leading order calculations. In this paper we have presented
an ${\cal O}(\alpha_s)$ calculation of the reaction
$p\,p\hskip-7pt\hbox{$^{^{(\!-\!)}}$} \rightarrow W^{\pm}Z + X
\rightarrow \ell_1^\pm \nu_1 \ell_2^+ \ell_2^- + X$ for general,
$C$ and $P$ conserving,
$WWZ$ couplings, using a combination of analytic and Monte Carlo
integration techniques. The leptonic decays $W\rightarrow \ell_1\nu_1$ and
$Z \rightarrow \ell_2^+ \ell_2^-$ have been included
in the narrow width approximation in our calculation.
Decay spin correlations are correctly taken into account in our
approach, except in the finite virtual contribution. The finite virtual
correction term contributes only at the few per cent level to the total
NLO cross section, thus decay spin correlations can be safely ignored here.

The $p_T^{}(Z)$ differential cross section
is very sensitive to non-standard $WWZ$ couplings. We found that
QCD corrections significantly change the shape of this distribution
at very high energies (see Fig.~\ref{FIG:PTLHC} and~\ref{FIG:KFAC}b).
This shape change is due to a combination of destructive
interference in the $WZ$ Born subprocess and a logarithmic enhancement
factor in the $qg$ and $\bar q g$ real emission subprocesses.
The destructive interference suppresses the size of the $WZ$ Born
cross section and is also responsible for the approximate amplitude zero
in $q_1\bar q_2\to WZ$~\cite{UJH}.
The logarithmic enhancement factor originates in the high $p_T^{}(Z)$
[$p_T^{}(W)$]
region of phase space where the $Z$ [$W$] boson is balanced by a
high $p_T^{}$ quark which radiates a soft $W$ [$Z$] boson. The
logarithmic enhancement factor and the large gluon density at high
center of mass energies
make the ${\cal O} (\alpha_s)$ corrections large for $p_T^{}(Z)\gg M_Z$.
Since the Feynman diagrams responsible for the enhancement at large
$p_T^{}(Z)$ do not involve the $WWZ$ vertex,
inclusive NLO QCD corrections to $W^\pm Z$ production tend to reduce the
sensitivity to non-standard couplings. QCD corrections in $WZ$
production thus exhibit the same features which characterize the ${\cal
O}(\alpha_s)$ corrections in $W\gamma$ production.

At the Tevatron and DiTevatron, $WZ$ production proceeds mainly via
quark-antiquark annihilation and, for the expected integrated
luminosities ($\leq 10$~fb$^{-1}$), large transverse momenta are not
accessible. As a result, the sensitivity reduction
in the high $p_T^{}$ tail caused by the QCD corrections is
balanced by the larger cross section at ${\cal O}(\alpha_s)$,
and the limits derived from the NLO and LO $p_T^{}(Z)$ distribution
are very similar (see Table~I). At the LHC, however, where the
$qg$ luminosity is very high and the change in slope of the SM $p_T^{}(Z)$
distribution from QCD corrections is very pronounced, the sensitivity
bounds which can be achieved are weakened by up to 40\% (see Table~II).

The size of the QCD corrections at large $p_T^{}(Z)$ may
be reduced substantially, and a fraction of the sensitivity to
anomalous $WWZ$ couplings which was lost at the LHC
may be regained, by imposing a jet veto, {\it i.e.}, by considering the
exclusive $WZ+0$~jet channel instead of inclusive $WZ+X$
production.  The improvement is equivalent to roughly a factor
1.5~--~2.5 increase in integrated luminosity. The
dependence of the NLO $WZ+0$~jet cross section on the factorization
scale $Q^2$ is significantly reduced compared to that of the inclusive
NLO $WZ+X$ cross section. Uncertainties which
originate from the variation of $Q^2$ will thus be smaller for
sensitivity bounds obtained from the $WZ+0$~jet channel than for those
derived from the inclusive NLO $WZ+X$ cross section. At the
Tevatron (DiTevatron), NLO QCD corrections do not
influence the sensitivity limits in a significant
way. Nevertheless, it will be important to take these corrections into
account when extracting information on the structure of the $WWZ$
vertex, in order to reduce systematic and theoretical errors.

At the Tevatron (DiTevatron) with $\int\!{\cal L}dt=10$~fb$^{-1}$,
taking into account all correlations between the different $WWZ$ couplings,
$\Delta\kappa^0$ can be measured with 70~--~100\% (50~--~60\%) accuracy
in $WZ$ production in the purely leptonic channels, whereas the two other
couplings can be determined with an uncertainty of 0.1~--~0.25. At the
LHC with
$\int\!{\cal L}dt=100$~fb$^{-1}$, $\Delta\kappa^0$ can be determined
with an uncertainty of about 10\%, whereas $\Delta g_1^0$ and
$\lambda^0$ can be measured to better than 0.01, with details depending
on the form factor scale assumed (see Table~II).

The bounds listed in Tables~I and~II should be compared with the limits
which can be obtained in other channels, and at $e^+e^-$ colliders.
Assuming SM $WW\gamma$ couplings, $\Delta g_1^0$ and
$\lambda^0$ can be measured in $p\bar p\to W^+W^-, W^\pm Z\to\ell^\pm\nu
jj$ with a precision similar to that which can be achieved in the $W^\pm
Z\to\ell_1^\pm\nu_1\ell_2^+\ell_2^-$ mode, both at the Tevatron with
1~fb$^{-1}$ and the TeV*. The limits which can be obtained for
$\Delta\kappa^0$ from the $\ell^\pm\nu jj$ final state
are about a factor~4 better than those from double leptonic $WZ$
decays~\cite{DPF}.
The bounds which can be achieved for $\Delta\kappa^0$, $\Delta g_1^0$
and $\lambda^0$ in $e^+e^-\to W^+W^-$ at LEP~II depend
quite sensitively on the center of mass energy. For
$\sqrt{s}=176$~GeV and $\int\!{\cal L}dt=500$~pb$^{-1}$, the $WWZ$
couplings can be measured with a precision of about $\pm 0.5$, if
correlations between the three couplings are taken into
account~\cite{DPF,BIL}. At a linear $e^+e^-$ collider with a center of mass
energy of 500~GeV or higher, they can be determined with an
accuracy of better than 0.01~\cite{DPF,BILL}.

We also studied possible experimental signals of the approximate zero in
the SM $WZ$ amplitude. Unlike the situation encountered in $W\gamma$
production where the radiation zero leads to a pronounced minimum in
the photon-lepton
rapidity difference distribution, the approximate amplitude zero in $WZ$
production causes a slight dip only in the corresponding $\Delta y(Z,
\ell_1)=y(Z)-y(\ell_1)$ distribution. In
$W^\pm\gamma$ production, the dominant $W$ helicity is $\lambda_W = \pm
1$, implying that the charged lepton from the decaying $W$ boson
tends to be
emitted in the direction of the parent $W$ boson, and thus reflects most of
its kinematic properties. In contrast, none of the $W$ helicities dominates
in $WZ$ production. The charged lepton originating from the $W$ boson
decay, $W\to\ell_1\nu_1$, thus only partly reflects the kinematic
properties of the parent $W$ boson, which reduces the significance of the
dip. At Tevatron and DiTevatron
energies, higher order QCD corrections only negligibly influence the
shape of the $\Delta y(Z,\ell_1)$ distribution. At the LHC, however, NLO
QCD effects completely obscure the dip, unless a 0-jet requirement is
imposed.

Alternatively, cross section ratios can be used to search for
experimental footprints of the approximate amplitude zero. We found that
the ratio of $ZZ$ to $WZ$ cross sections as a function of the minimum
$Z$ boson transverse momentum, $p_T^{\rm min}$, increases with $p_T^{\rm
min}$ for values larger than 100~GeV. The increase of the ratio of $ZZ$
to $WZ$ cross sections is a direct consequence of the approximate
zero. The ratio of $WZ$ to $W\gamma$ cross sections, on the other hand,
is almost independent of the minimum $p_T^{}$ of the $Z$ boson and photon
for sufficiently large values of $p_T^{\rm min}$,
indicating that the approximate zero in $WZ$ production and
the radiation zero in $W\gamma$ production affect the $Z$ boson and photon
transverse momentum distributions in a very similar way. QCD corrections
have a significant impact on the $ZZ$ to $WZ$ cross section ratio at
the LHC unless a jet veto is imposed, whereas they largely cancel in
the $WZ$ to $W\gamma$ cross section ratio.

Together with the $\Delta y(Z,\ell_1)$ distribution, the $ZZ$ to $W^\pm
Z$ and $WZ$ to $W\gamma$ cross section ratios are useful tools in
searching for
the approximate amplitude zero in $WZ$ production. However, for
the integrated luminosities envisioned, it will not be easy
to conclusively establish the approximate amplitude zero in $WZ$
production at the Tevatron or the LHC.

\newpage
%%%%%%%%%%%%%%%%%%%%%%%%% ACKNOWLEDGMENTS %%%%%%%%%%%%%%%%%%%%%%%%%%%%%%%%
%
\acknowledgements

We would like to thank B.~Choudhury, T.~Fuess, C.~Wendt, and D.~Zeppenfeld
for stimulating discussions. Two of us (U.B. and T.H.) wish to thank the
Fermilab Theory Group for its warm hospitality during various stages of
this work. This work has been supported in part by Department of Energy
grants Nos.~DE-FG03-91ER40674 and DE-FG05-87ER40319
and by Texas National Research Laboratory grant No.~RGFY93-330.
T.H. is also supported in part by a UC-Davis Faculty Research Grant.

%
%%%%%%%%%%%%%%%%%%%%% REFERENCES %%%%%%%%%%%%%%%%%%%%%%%%%%%%%%%%%%%%%%%%%
%

%
\newpage
%
%%%%%%%%%%%%%%%%%%%%%%%%% TABLES %%%%%%%%%%%%%%%%%%%%%%%%%%%%%%%%%%%%%%%%
\widetext
\newcommand{\crc}{\crcr\noalign{\vskip -8pt}}
\begin{table}
\caption{Sensitivities achievable at the 95\% confidence
level (CL) for the anomalous $WWZ$ couplings $\Delta g_1^0$,
$\Delta\kappa^0$, and
$\lambda^0$ in $p\bar p\rightarrow W^\pm Z + X\rightarrow
\ell_1^\pm \nu_1 \ell_2^+ \ell_2^-+X$ at a) leading order and b)
next-to-leading order for
the Tevatron, the TeV* \protect{(}$\protect{\sqrt{s}=1.8}$~TeV in both
cases\protect{)}, and the DiTevatron
\protect{(}$\protect{\sqrt{s}=3.5}$~TeV\protect{)}. The limits for
each coupling apply for arbitrary values of the two other couplings.
For the form factors we use Eqs.~(\protect{\ref{EQ:GONEFORM}}),
(\protect{\ref{EQ:KAPPAFORM}}), and~(\protect{\ref{EQ:LAMBDAFORM}})
with $n=2$ and $\Lambda_{FF}=1$~TeV. The cuts summarized in
Sec.~IIIB are imposed. \protect{\\ [3.mm]} }
\label{TABLE1}
\begin{tabular}{cccc}
\multicolumn{4}{c}{a) leading order}\\
\multicolumn{1}{c}{ }
&\multicolumn{1}{c}{Tevatron}
&\multicolumn{1}{c}{TeV*}
&\multicolumn{1}{c}{DiTevatron}\\
\multicolumn{1}{c}{coupling}
&\multicolumn{1}{c}{$\int\!{\cal L}dt=1$~fb$^{-1}$}
&\multicolumn{1}{c}{$\int\!{\cal L}dt=10$~fb$^{-1}$}
&\multicolumn{1}{c}{$\int\!{\cal L}dt=10$~fb$^{-1}$}\\
\tableline
 & & & \\ [-6.mm]
$\Delta g_1^0$ & $\matrix{+0.52 \crc -0.29}$ & $\matrix{+0.26 \crc -0.
10}$ & $\matrix{+0.15 \crc -0.05}$ \\ [5.mm]
$\Delta\kappa^0$ & $\matrix{+1.9 \crc -1.4}$ & $\matrix{+1.0
\crc -0.7}$ & $\matrix{+0.5 \crc -0.5}$ \\ [5.mm]
$\lambda^0$ & $\matrix{+0.34 \crc -0.37}$ & $\matrix{+0.15
\crc -0.15}$ & $\matrix{+0.08 \crc -0.07}$ \\ [5.mm]
\tableline
\tableline
\multicolumn{4}{c}{b) next-to-leading order}\\
\multicolumn{1}{c}{ }
&\multicolumn{1}{c}{Tevatron}
&\multicolumn{1}{c}{TeV*}
&\multicolumn{1}{c}{DiTevatron}\\
\multicolumn{1}{c}{coupling}
&\multicolumn{1}{c}{$\int\!{\cal L}dt=1$~fb$^{-1}$}
&\multicolumn{1}{c}{$\int\!{\cal L}dt=10$~fb$^{-1}$}
&\multicolumn{1}{c}{$\int\!{\cal L}dt=10$~fb$^{-1}$}\\
\tableline
 & & & \\ [-6.mm]
$\Delta g_1^0$ & $\matrix{+0.50 \crc -0.27}$ & $\matrix{+0.25 \crc -0.
10}$ & $\matrix{+0.15 \crc -0.05}$ \\ [5.mm]
$\Delta\kappa^0$ & $\matrix{+1.8 \crc -1.2}$ & $\matrix{+1.0
\crc -0.7}$ & $\matrix{+0.6 \crc -0.5}$ \\ [5.mm]
$\lambda^0$ & $\matrix{+0.33 \crc -0.34}$ & $\matrix{+0.15
\crc -0.14}$ & $\matrix{+0.08 \crc -0.08}$ \\ [5.mm]
\end{tabular}
\end{table}
\newpage
\begin{table}
\caption{Sensitivities achievable at the 95\% confidence
level (CL) for the anomalous $WWZ$ couplings $\Delta g_1^0$,
$\Delta\kappa^0$, and $\lambda^0$ in $pp\rightarrow W^+ Z + X\rightarrow
\ell_1^+ \nu_1 \ell_2^+ \ell_2^-+X$ at the LHC
\protect{(}$\protect{\sqrt{s}=14}$~TeV\protect{)}. The limits for
each coupling apply for arbitrary values of the two other couplings.
For the form factors we use Eqs.~(\protect{\ref{EQ:GONEFORM}}),
(\protect{\ref{EQ:KAPPAFORM}}), and~(\protect{\ref{EQ:LAMBDAFORM}})
with $n=2$. The cuts summarized in Sec.~IIIB are imposed. In the NLO
0-jet case we have used the jet definition of
Eq.~(\protect{\ref{EQ:SSCJET}}).
\protect{\\ [-2.mm]} }
\label{TABLE2}
\begin{tabular}{cccc}
\multicolumn{4}{c}{a) $\int\!{\cal L}dt=10$~fb$^{-1}$,
$\Lambda_{FF}=1$~TeV}\\
\multicolumn{1}{c}{coupling}
&\multicolumn{1}{c}{Born appr.}
&\multicolumn{1}{c}{incl. NLO}
&\multicolumn{1}{c}{NLO 0-jet}\\
\tableline
 & & & \\ [-6.mm]
$\Delta g_1^0$ & $\matrix{+0.108 \crc -0.036}$ & $\matrix{+0.127 \crc
-0.044}$ & $\matrix{+0.115 \crc -0.038}$ \\ [5.mm]
$\Delta\kappa^0$ & $\matrix{+0.53 \crc -0.46}$ & $\matrix{+0.62 \crc
-0.59}$ & $\matrix{+0.60 \crc -0.49}$ \\ [5.mm]
$\lambda^0$ & $\matrix{+0.058 \crc -0.057}$ & $\matrix{+0.065 \crc
-0.073}$ & $\matrix{+0.063 \crc -0.062}$ \\ [5.mm]
\tableline
\tableline
\multicolumn{4}{c}{b) $\int\!{\cal L}dt=100$~fb$^{-1}$,
$\Lambda_{FF}=1$~TeV}\\
\multicolumn{1}{c}{coupling}
&\multicolumn{1}{c}{Born appr.}
&\multicolumn{1}{c}{incl. NLO}
&\multicolumn{1}{c}{NLO 0-jet}\\
\tableline
 & & & \\ [-6.mm]
$\Delta g_1^0$ & $\matrix{+0.082 \crc -0.014}$ & $\matrix{+0.096 \crc
-0.018}$ & $\matrix{+0.085 \crc -0.016}$ \\ [5.mm]
$\Delta\kappa^0$ & $\matrix{+0.17 \crc -0.34}$ & $\matrix{+0.24 \crc
-0.41}$ & $\matrix{+0.20 \crc -0.35}$ \\ [5.mm]
$\lambda^0$ & $\matrix{+0.038 \crc -0.036}$ & $\matrix{+0.042 \crc
-0.048}$ & $\matrix{+0.036 \crc -0.038}$ \\ [5.mm]
\tableline
\tableline
\multicolumn{4}{c}{c) $\int\!{\cal L}dt=100$~fb$^{-1}$,
$\Lambda_{FF}=3$~TeV}\\
\multicolumn{1}{c}{coupling}
&\multicolumn{1}{c}{Born appr.}
&\multicolumn{1}{c}{incl. NLO}
&\multicolumn{1}{c}{NLO 0-jet}\\
\tableline
 & & & \\ [-6.mm]
$\Delta g_1^0$ & $\matrix{+0.0164 \crc -0.0048}$ & $\matrix{+0.0200
\crc -0.0066}$ & $\matrix{+0.0188 \crc -0.0050}$ \\ [5.mm]
$\Delta\kappa^0$ & $\matrix{+0.092 \crc -0.120}$ & $\matrix{+0.128
\crc -0.160}$ & $\matrix{+0.108 \crc -0.132}$ \\ [5.mm]
$\lambda^0$ & $\matrix{+0.0084 \crc -0.0082}$ & $\matrix{+0.0102
\crc -0.0100}$ & $\matrix{+0.0092 \crc -0.0090}$ \\ [5.mm]
\end{tabular}
\end{table}
\newpage
%
%%%%%%%%%%%%%%%%%%%%%% FIGURE CAPTIONS %%%%%%%%%%%%%%%%%%%%%%%%%%%%%%%%%%
%
% FIG. 1
\begin{figure}
\caption{Feynman rule for the general $WWZ$ vertex.
The factor $g_{WWZ}=e\cot\theta_{\rm W}$ is the $WWZ$ coupling strength
and $Q_W$ is the electric charge of the $W$ boson. The vertex function
$\Gamma_{\beta \mu \nu}(k,k_1,k_2)$ is given in
Eq.~(\protect{\ref{EQ:NSMCOUPLINGS}}).}
\label{FIG:VERTEX}
\end{figure}
%
% FIG. 2
\begin{figure}
\caption{The inclusive differential cross section for the reconstructed
$WZ$ mass in the reaction
$\protect{p \bar p \to W^+Z + X \to\ell_1^+\nu_1\ell_2^+\ell_2^- + X}$
at $\protect{\sqrt{s} = 1.8}$~TeV; a) in the Born approximation and b)
including NLO QCD corrections.
The curves are for the SM (solid lines), $\lambda^0 = -0.5$ (dashed
lines), $\Delta\kappa^0 = -1.0$ (dotted lines), and $\Delta g_1^0 =
-0.5$ (dot-dashed lines). The cuts imposed are summarized in Sec.~IIIB. }
\label{FIG:MWZTEV}
\end{figure}
%
% FIG. 3
\begin{figure}
\caption{The inclusive differential cross section for the reconstructed
$WZ$ mass in the reaction
$\protect{p \bar p \to W^+Z + X \to\ell_1^+\nu_1\ell_2^+\ell_2^- + X}$
at $\protect{\sqrt{s} = 3.5}$~TeV; a) in the Born approximation and b)
including NLO QCD corrections.
The curves are for the SM (solid lines), $\lambda^0 = -0.5$ (dashed
lines), $\Delta\kappa^0 = -1.0$ (dotted lines), and $\Delta g_1^0 =
-0.5$ (dot-dashed lines). The cuts imposed are summarized in Sec.~IIIB. }
\label{FIG:MWZDITEV}
\end{figure}
%
% FIG. 4
\begin{figure}
\caption{The inclusive differential cross section for the reconstructed
$WZ$ mass in the reaction
$\protect{pp \to W^+Z + X \to\ell_1^+\nu_1\ell_2^+\ell_2^- + X}$
at $\protect{\sqrt{s} = 14}$~TeV; a) in the Born approximation and b)
including NLO QCD corrections.
The curves are for the SM (solid lines), $\lambda^0 = -0.25$ (dashed
lines), $\Delta\kappa^0 = -1.0$ (dotted lines), and $\Delta g_1^0 =
-0.25$ (dot-dashed lines). The cuts imposed are summarized in Sec.~IIIB. }
\label{FIG:MWZLHC}
\end{figure}
%
% FIG. 5
\begin{figure}
\caption{\sloppy{
The inclusive NLO differential cross section for the reconstructed
$WZ$ mass in the reaction a)
$\protect{p\bar p \to W^+Z + X \to\ell_1^+\nu_1\ell_2^+\ell_2^- + X}$
at $\protect{\sqrt{s} = 1.8}$~TeV and b)
$\protect{pp \to W^+Z + X}$ $\protect{\to\ell_1^+\nu_1\ell_2^+\ell_2^-
+ X}$ at $\protect{\sqrt{s} = 14}$~TeV.
The curves are for the SM with reconstructed invariant mass (solid
lines) and non-standard $WWZ$ couplings (dotted, dashed, and
dot-dashed curves) as listed on the figure.
The lower (upper) lines apply for positive
(negative) anomalous couplings. The dash-double-dotted line shows the
true SM $WZ$ invariant mass distribution.
The cuts imposed are summarized in Sec.~IIIB.} }
\label{FIG:MWZ}
\end{figure}
%
% FIG. 6
\begin{figure}
\caption{The inclusive NLO differential cross section for the cluster
transverse mass for a)
$\protect{p \bar p \to W^+Z + X \to\ell_1^+\nu_1\ell_2^+\ell_2^- + X}$
at $\protect{\sqrt{s} = 1.8}$~TeV and b)
$\protect{pp \to W^+Z + X \to\ell_1^+\nu_1\ell_2^+\ell_2^- + X}$
at $\protect{\sqrt{s} = 14}$~TeV. The curves are for the SM (solid
lines) and non-standard $WWZ$ couplings (dotted, dashed, and
dot-dashed curves) as listed on the figure.
The cuts imposed are summarized in Sec.~IIIB. }
\label{FIG:MTCL}
\end{figure}
%
% FIG. 7
\begin{figure}
\caption{The inclusive differential cross section for the transverse
momentum of the $Z$ boson in the reaction
$\protect{p \bar p \to W^+Z + X \to\ell_1^+\nu_1\ell_2^+\ell_2^- + X}$
at $\protect{\sqrt{s} = 1.8}$~TeV; a) in the Born approximation and b)
including NLO QCD corrections.
The curves are for the SM (solid lines), $\lambda^0 = -0.5$ (dashed
lines), $\Delta\kappa^0 = -1.0$ (dotted lines), and $\Delta g_1^0 =
-0.5$ (dot-dashed lines). The cuts imposed are summarized in Sec.~IIIB. }
\label{FIG:PTTEV}
\end{figure}
%
% FIG. 8
\begin{figure}
\caption{The inclusive differential cross section for the transverse
momentum of the $Z$ boson in the reaction
$\protect{p \bar p \to W^+Z + X \to\ell_1^+\nu_1\ell_2^+\ell_2^- + X}$
at $\protect{\sqrt{s} = 3.5}$~TeV; a) in the Born approximation and b)
including NLO QCD corrections.
The curves are for the SM (solid lines), $\lambda^0 = -0.5$ (dashed
lines), $\Delta\kappa^0 = -1.0$ (dotted lines), and $\Delta g_1^0 =
-0.5$ (dot-dashed lines). The cuts imposed are summarized in Sec.~IIIB. }
\label{FIG:PTDITEV}
\end{figure}
%
% FIG. 9
\begin{figure}
\caption{The inclusive differential cross section for the
transverse momentum of the $Z$ boson in the reaction
$\protect{pp \to W^+Z + X \to\ell_1^+\nu_1\ell_2^+\ell_2^- + X}$
at $\protect{\sqrt{s} = 14}$~TeV; a) in the Born approximation and b)
including NLO QCD corrections.
The curves are for the SM (solid lines), $\lambda^0 = -0.25$ (dashed
lines), $\Delta\kappa^0 = -1.0$ (dotted lines), and $\Delta g_1^0 =
-0.25$ (dot-dashed lines). The cuts imposed are summarized in Sec.~IIIB. }
\label{FIG:PTLHC}
\end{figure}
%
% FIG. 10
\begin{figure}
\caption{Ratio of the NLO to LO differential cross sections of the $Z$
boson transverse momentum as a function of $p_T^{}(Z)$ for a)
$\protect{p \bar p \to W^+Z + X \to\ell_1^+\nu_1\ell_2^+\ell_2^- + X}$
and b) $\protect{pp \to W^+Z + X \to\ell_1^+\nu_1\ell_2^+\ell_2^- + X}$
at $\protect{\sqrt{s} = 14}$~TeV. In part a) the solid (dotted) and
dashed (dot-dashed) lines give the ratio in the SM and for
$\lambda^0=-0.5$ at the Tevatron (DiTevatron), respectively. In part b)
the solid and dashed curves show the ratio for the SM and $\lambda^0=-0.
25$ at the LHC. The cuts imposed are summarized in Sec.~IIIB. }
\label{FIG:KFAC}
\end{figure}
%
% FIG. 11
\begin{figure}
\caption{The differential cross section for the $Z$ boson transverse
momentum in the reaction
$\protect{p\bar p \to W^+Z + X \to\ell_1^+\nu_1\ell_2^+\ell_2^- + X}$
at $\protect{\sqrt{s} = 1.8}$~TeV in the SM.
a) The inclusive NLO differential cross section (solid line) is
shown, together with the \protect{${\cal O}(\alpha_s)$} \protect{0-jet}
(dotted line)
and the (LO) \protect{1-jet} (dot dashed line) exclusive differential
cross sections, using the jet definition in
Eq.~(\protect{\ref{EQ:TEVJET}}).
b) The NLO $\protect{WZ +0}$~jet exclusive differential cross section
(dotted line) is compared with the Born differential cross section
(dashed line). The cuts imposed are summarized in Sec.~IIIB. }
\label{FIG:PTZTEVEX}
\end{figure}
%
% FIG. 12
\begin{figure}
\caption{The differential cross section for the $Z$ boson transverse
momentum in the reaction
$\protect{pp \to W^+Z + X \to\ell_1^+\nu_1\ell_2^+\ell_2^- + X}$
at $\protect{\sqrt{s} = 14}$~TeV in the SM.
a) The inclusive NLO differential cross section (solid line) is
shown, together with the \protect{${\cal O}(\alpha_s)$} \protect{0-jet}
(dotted line)
and the (LO) \protect{1-jet} (dot dashed line) exclusive differential
cross sections, using the jet definition in
Eq.~(\protect{\ref{EQ:SSCJET}}).
b) The NLO $\protect{WZ +0}$~jet exclusive differential cross section
(dotted line) is compared with the Born differential cross section
(dashed line). The cuts imposed are summarized in Sec.~IIIB. }
\label{FIG:PTZLHCEX}
\end{figure}
%
% FIG. 13
\begin{figure}
\caption{The total cross section for \protect{
$p\,p\hskip-7pt\hbox{$^{^{(\!-\!)}}$} \rightarrow W^+ Z + X
\rightarrow \ell_1^+\nu_1\ell_2^+\ell_2^- + X$} in the SM
versus the scale $Q$; a) at the Tevatron and b) at the
LHC. The curves represent the inclusive NLO (solid lines), the Born
(dot-dashed lines), the LO \protect{1-jet} exclusive (dashed lines),
and the NLO \protect{0-jet} exclusive (dotted lines) cross sections.
The cuts imposed are summarized in Sec.~IIIB. For the jet definitions,
we have used Eqs.~(\protect{\ref{EQ:TEVJET}})
and~(\protect{\ref{EQ:SSCJET}}). }
\label{FIG:QSCALE}
\end{figure}
%
% FIG. 14
\begin{figure}
\caption{Limit contours at the \protect{95\%~CL} for
\protect{$p\bar p\rightarrow
W^\pm Z+X\rightarrow\ell_1^\pm\nu_1\ell_2^+\ell_2^-+X$} derived from the
inclusive NLO $p_T^{}(Z)$ distribution.
Contours are shown in three planes:
a) the \protect{$\Delta\kappa^0$} \protect{--}~\protect{$\lambda^0$} plane,
b) the \protect{$\Delta\kappa^0$} \protect{--}~\protect{$\Delta g_1^0$}
plane,
and c) the \protect{$\Delta g_1^0$} \protect{--}~\protect{$\lambda^0$}
plane. The solid and dashed lines give the results for the Tevatron
($\protect{\sqrt{s}=1.8}$~TeV) with
\protect{$\int\!{\cal L}dt= 1$}~fb\protect{$^{-1}$} and
\protect{$\int\!{\cal L}dt=10$}~fb\protect{$^{-1}$},
respectively. The dotted curve shows the result obtained for the
DiTevatron ($\protect{\sqrt{s}=3.5}$~TeV)
with \protect{$\int\!{\cal L}dt=10$}~fb\protect{$^{-1}$}.
The cuts imposed are summarized in Sec.~IIIB.}
\label{FIG:PPBARLIM}
\end{figure}
%
% FIG. 15
\begin{figure}
\caption{Limit contours at the \protect{95\%~CL} for
\protect{$pp\rightarrow W^+
Z+X\rightarrow\ell_1^+\nu_1\ell_2^+\ell_2^-+X$} at
$\protect{\sqrt{s}=14}$~TeV derived from the
\protect{$p_T^{}(Z)$} distribution. Contours are shown in three planes:
a) the \protect{$\Delta\kappa^0$}\protect{--}~\protect{$\lambda^0$} plane,
b) the \protect{$\Delta\kappa^0$}\protect{--}~\protect{$\Delta g_1^0$} plane,
and c) the \protect{$\Delta g_1^0$}\protect{--}~\protect{$\lambda^0$} plane.
The solid and dashed lines give the inclusive NLO and LO results,
respectively, for \protect{$\int\!{\cal L}dt=10$}~fb\protect{$^{-1}$}.
The dotted and dot-dashed curves show the results obtained from the
exclusive NLO $WZ+0$~jet channel for integrated luminosities of 10~fb$^{-1}$
and 100~fb$^{-1}$, respectively. The cuts imposed are summarized
in Sec.~IIIB.}
\label{FIG:PPLIM}
\end{figure}
%
% FIG. 16
\begin{figure}
\caption{Rapidity spectrum of the $Z$ boson in the $WZ$ rest frame in
the Born approximation for a)
$\protect{p\bar p \to W^+Z + X \to\ell_1^+\nu_1\ell_2^+\ell_2^- + X}$
at $\protect{\sqrt{s} = 1.8}$~TeV and b)
$\protect{pp \to W^+Z + X \to\ell_1^+\nu_1\ell_2^+\ell_2^- + X}$
at $\protect{\sqrt{s} = 14}$~TeV.
The curves are for the SM with reconstructed $WZ$ rest frame (solid
lines) and non-standard $WWZ$ couplings (dotted, dashed, and
dot-dashed curves) as listed on the figure.
The lower (upper) lines apply for positive
(negative) anomalous couplings. The dash-double-dotted line shows the
true SM $|y^*(Z)|$ distribution. The cuts imposed are summarized in
Sec.~IIIB. }
\label{FIG:YSTAR}
\end{figure}
%
% FIG. 17
\begin{figure}
\caption{\sloppy{SM rapidity difference distributions
in the Born approximation for a) $\protect{p\bar p \to}$ $\protect{W^+Z
+ X \to\ell_1^+ \nu_1\ell_2^+\ell_2^- + X}$ at $\protect{\sqrt{s} =
1.8}$~TeV and b) $\protect{pp \to W^+Z + X \to}$
$\protect{\ell_1^+\nu_1\ell_2^+\ell_2^- + X}$ at $\protect{\sqrt{s}
= 14}$~TeV. The curves are for $y(Z)-y(\ell_1^+)$
(solid lines) and $y(\ell_2^-)-y(\ell_1^+)$ (dotted lines).
The cuts imposed are summarized in Sec.~IIIB.} }
\label{FIG:RAPDIFF}
\end{figure}
%
% FIG. 18
\begin{figure}
\caption{The differential cross section for the
rapidity difference $\Delta y(Z,\ell_1)$ for
$\protect{p\bar p \to}$ $\protect{ W^+Z + X \to\ell_1^+\nu_1\ell_2^+
\ell_2^- + X}$ a)
at $\protect{\sqrt{s} = 1.8}$~TeV and b) at $\protect{\sqrt{s} = 3.
5}$~TeV. The solid and dot-dashed curves show the inclusive NLO and the
LO SM prediction, respectively. The dashed and dotted lines give the
results for $\Delta\kappa^0=+1$ and $\Delta\kappa^0=-1$, respectively.
The error bars
associated with the solid curves indicate the expected statistical
uncertainties for an integrated luminosity of 10~fb$^{-1}$. The cuts
imposed are summarized in Sec.~IIIB.}
\label{FIG:DELYPPBAR}
\end{figure}
%
% FIG. 19
\begin{figure}
\caption{The differential cross section for the rapidity difference
$\Delta y(Z,\ell_1)$ at $\protect{\sqrt{s} = 14}$~TeV for a)
$\protect{pp \to}$ $\protect{ W^+Z + X \to\ell_1^+\nu_1\ell_2^+\ell_2^-
+ X}$ in the SM and b) for $\protect{pp \to W^+Z + 0}$~jet
$\protect{\to\ell_1^+\nu_1\ell_2^+\ell_2^- + 0}$~jet at NLO. In part a)
the dotted and dashed curves show the inclusive NLO and the
LO SM prediction, respectively, while the solid curve
gives the prediction for the SM NLO $WZ+0$~jet case. The error bars
associated with the solid curve indicate the expected statistical
uncertainties for an integrated luminosity of 100~fb$^{-1}$.
In part b) the curves are for the SM (solid line),
$\Delta\kappa^0=+1$ (dotted line), $\lambda^0=+0.25$ (dashed curve), and
$\Delta g_1^0=+0.25$ (dot-dashed curve).
The cuts imposed are summarized in Sec.~IIIB. For the jet definition,
we have used Eq.~(\protect{\ref{EQ:SSCJET}}).}
\label{FIG:DELYPP}
\end{figure}
%
% FIG. 20
\begin{figure}
\caption{The ratio ${\cal R}_{ZZ/WZ}=B(Z\to\ell^+\ell^-)\,\sigma(ZZ)/
B(W\to\ell\nu)\,\sigma(W^\pm Z)$, $\ell=e,\,\mu$, as a function of the
minimum transverse
momentum of the $Z$ boson, $p_T^{\rm min}$, at a) the Tevatron and b)
the LHC. The solid and dashed line show the inclusive NLO and the LO
result for the SM, respectively. The dotted line in b) gives the SM
cross section ratio at NLO if a 0-jet requirement is imposed. The
dot-dashed line
displays ${\cal R}_{ZZ/WZ}$ for $\Delta\kappa^0=+1$. In part a) this
curve is calculated taking into account inclusive NLO QCD corrections,
whereas in part b) the dot-dashed curve is for the NLO 0-jet cross section
ratio. The cuts imposed are summarized in Sec.~IIIB. For the jet
definition, we have used Eq.~(\protect{\ref{EQ:SSCJET}}).}
\label{FIG:RZZ}
\end{figure}
%
% FIG. 21
\begin{figure}
\caption{The ratio ${\cal R}_{WZ/W\gamma}=B(Z\to\ell^+\ell^-)\,
\sigma(W^\pm Z)/\sigma(W^\pm\gamma)$, $\ell=e,\,\mu$, as a function of the
minimum transverse
momentum of the $Z$ boson and photon, $p_T^{\rm min}$, respectively, at
a) the Tevatron and b)
the LHC. The solid and dashed lines show the inclusive NLO and the LO
result for the SM, respectively. The dotted and dot-dashed lines
display the inclusive NLO and LO $WZ$ to $W\gamma$ cross section ratio for
$\Delta\kappa_\gamma^0=\Delta\kappa^0=-1$. Here, $\Delta\kappa_\gamma$
is the anomalous $WW\gamma$ coupling defined in an analogous way to
$\Delta\kappa$ [see Eq.~(\protect{\ref{EQ:LAGRANGE}})]. The cuts imposed are
summarized in Sec.~IIIB and Eq.~(\protect{\ref{EQ:WGCUT}}).}
\label{FIG:RWG}
\end{figure}
%
%%%%%%%%%%%%%%%%%%%%%%%%%%%%%%%%%%%%%%%%%%%%%%%%%%%%%%%%%%%%%%%%%%%%%%%%%%%%%
%
\end{document}